\documentclass[format=acmsmall, review=false, screen=true, acmtsc]{acmart}

\usepackage{multirow}
\usepackage{amsmath}
\usepackage{amssymb}
\usepackage{type1cm}     % type1 computer modern font
\usepackage{graphicx}     % advanced figures
\usepackage{xspace}     % fix space in macros
\usepackage{balance}     % to better equalize the last page
\usepackage{booktabs}     % nicer tables
\PassOptionsToPackage{hyphens}{url}
\usepackage[hyphens]{url}     % handle long urls
\usepackage{siunitx}          % \num for decimal grouping
\usepackage[vlined,linesnumbered,ruled,noend]{algorithm2e}     % algorithms
\usepackage[font={small}]{subfig} % subfig, 4 figures in a row
\usepackage{epstopdf}

\usepackage[font=small,labelfont=bf,tableposition=top]{caption} 

\usepackage[show]{chato-notes}
\graphicspath{{img/}}

\usepackage{pifont}
\newcommand{\cmark}{\ding{51}}
\newcommand{\xmark}{\ding{55}}

\newcommand{\spara}[1]{\smallskip\noindent\textbf{#1}}

\newenvironment {squishlist}
{\begin{list}{$\bullet$}
  { \setlength{\itemsep}{1pt}
     \setlength{\parsep}{1pt}
     \setlength{\topsep}{1pt}
     \setlength{\partopsep}{1pt}
     \setlength{\leftmargin}{1.5em}
     \setlength{\labelwidth}{1em}
     \setlength{\labelsep}{0.5em} } }
{\end{list}}

\newcommand{\rwc}{\ensuremath{\mathit{RWC}}\xspace}
\newcommand{\similar}{\ensuremath{\mathit{sim}}}

\newcommand{\prob}{\ensuremath{\mathit{Pr}}}
\newcommand{\bc}{\ensuremath{\mathit{bc}}}

\begin{document}

\setcopyright{acmlicensed}
\acmJournal{TSC}
\acmYear{2017} \acmVolume{1} \acmNumber{1} \acmArticle{1} \acmMonth{1} \acmPrice{\$15.00}\acmDOI{10.1145/3140565}

\title{Quantifying Controversy on Social Media}
\author{Kiran Garimella}
\affiliation{%
  \institution{Aalto University}
  \city{Helsinki} 
  \country{Finland} 
}
\email{kiran.garimella@aalto.fi}
% \authornote{Contact authors.}

\author{Gianmarco De~Francisci~Morales}
\affiliation{%
  \institution{Qatar Computing Research Institute}
  \city{Doha} 
  \country{Qatar} 
}
\email{gdfm@acm.org}

\author{Aristides Gionis}
\affiliation{%
  \institution{Aalto University}
  \city{Helsinki} 
  \country{Finland} 
}
\email{aristides.gionis@aalto.fi}

\author{Michael Mathioudakis}
\affiliation{%
  \institution{CNRS LIRIS \& INSA Lyon}
  \city{Lyon} 
  \country{France} 
}
\email{michael.mathioudakis@liris.cnrs.fr}

\renewcommand{\shortauthors}{K. Garimella, G. De Francisci Morales, A. Gionis, M. Mathioudakis}

\begin{abstract}
\emph{Which topics spark the most heated debates on social media?}
Identifying those topics is not only interesting from a societal point of view, 
but also allows the filtering and aggregation of social media content
for disseminating news stories.
% Identifying such topics is a first step towards creating systems that pierce echo chambers.
In this paper, we perform a systematic methodological study of controversy detection 
by using the content and the network structure of social media.

Unlike previous work, rather than study controversy in a single hand-picked topic and use domain-specific knowledge, we take a general approach to study topics \emph{in any domain}. 
Our approach to quantifying controversy is based on a graph-based three-stage pipeline, which involves 
($i$) building a \emph{conversation graph} about a topic; 
% which represents alignment of opinion among users; 
($ii$) partitioning the conversation graph to identify potential sides of the controversy; and 
($iii$)~measuring the amount of controversy from characteristics of the graph.

We perform an extensive comparison of controversy measures, 
different graph-building approaches, and data sources.
We use both controversial and non-controversial topics on Twitter, as well as other external datasets.
We find that our new random-walk-based measure outperforms existing ones in capturing the intuitive notion of controversy, and show that content features are vastly less helpful in this task.
\end{abstract}

% ACM classification



\maketitle

%\note[AG]{I changed the second sentence of the intro.
%The previous sentence was ``Identifying these topics is a first step towards creating systems that pierce echo chambers.'' However, the step from controversial topics to piercing echo chambers is not so obvious 
%(at least not obvious enough for being the first claim)
%neither is so central to our work.}

%%%%%%%%%%%%%%%%%%%%%%%%%%%%%%%%%%%%%%%%%%%%%%%%%%%%%%%%%%%%%%%%%%
%introduction

\section{Introduction}
\label{sec:intro}
\enlargethispage{0.8\baselineskip}

% explain: social media
Given their widespread diffusion, online social media have become increasingly important in the study of social phenomena such as peer influence, framing, bias, and controversy.
Ultimately, we would like to understand how users perceive the world through the lens of their social media feed.
%For instance, to offer users the possibility to balance their ``news diet'' \cite{kulshrestha2015characterizing,lacour2012balanced} on controversial topics by recommending contrarian content, which supports a view that differs from what they are mostly exposed to~\cite{munson2013encouraging}.
However, before addressing these advanced application scenarios, we first need to focus on the fundamental yet challenging task of distinguishing whether a topic of discussion is controversial.
%which represents a first fundamental step before confronting with even more ambitious tasks.
Our work is motivated by interest in observing controversies at societal level, monitoring their evolution, and possibly understanding which issues become controversial and why.

% Online social media have recently emerged as a new venue for public debate on current issues.
% Through social media, people not only get informed about breaking news and current events, but can also engage in discussions and  express their opinion on issues of interest.
% Public debates that appear on social media are typically spontaneous, uncoordinated, and easy for anyone to join.
% Topics of discussion might concern long-standing issues, but often also emerge from breaking news and current events,  and are shaped by public interest and grassroots dynamics.
% These characteristics, coupled with the fact that the majority of the content is open and easily accessible for analysis, make social media a large-scale observatory for studying controversy.

%In this paper, we focus on the topic of controversy, and we seek ways to identify issues that are controversial in social media.
The study of controversy in social media is not new;
there are many previous studies aimed at identifying and characterizing controversial issues, 
mostly around
political debates~\cite{adamic2005political,conover2011political,mejova2014controversy,morales2015measuring}
but also for other topics~\cite{guerra2013measure}.
And while most recent papers have focused on 
Twitter~\cite{conover2011political,guerra2013measure,mejova2014controversy,morales2015measuring}, 
controversy in other platforms, 
such as blogs~\cite{adamic2005political} and
opinion fora~\cite{akoglu2014quantifying},
has also been analyzed. 

However, most previous papers have severe limitations. 
First, the majority of previous studies focus on controversy regarding political issues, 
and, in particular, are centered around long-lasting major events, 
such as elections~\cite{adamic2005political, conover2011political}.
More crucially, most previous works can be characterized as \emph{case studies}, 
where controversy is identified in a single carefully-curated dataset, 
collected using ample domain knowledge and auxiliary domain-specific sources 
(e.g., an extensive list of hashtags regarding a major political event, or a list of left-leaning and right-leaning blogs).

We aim to overcome these limitations. 
We develop a framework to identify controversy regarding topics in any domain
(e.g., political, economical, or cultural), 
and without prior domain-specific knowledge about the topics in question.
Within the framework, we quantify the controversy associated with each topic, and thus compare different topics in order to find the most controversial ones.
Having a framework with these properties allows us to deploy a system in-the-wild, and is valuable for building real-world applications.

In order to enable such a versatile framework, 
we work with topics that are defined in a lightweight and domain-agnostic manner. 
Specifically, when focusing on Twitter, 
a topic is specified as a text query. %  and the related activity consisting of posts that match the query.
For example, ``\#beefban'' is a special keyword (a ``hashtag'') that
Twitter users employed in March 2015 to signal that their posts
referred to a decision by the Indian government about
the consumption of beef meat in India. 
In this case, the query ``\#beefban'' defines a topic of discussion, 
and the related activity consists of all posts that contain the query,
or other closely related terms and hashtags,
as explained in Section~\ref{sec:hashtags_to_clusters}.

% We represent a topic of discussion with a
% \emph{conversation graph}:
% in such a graph, vertices represent people, 
% and edges represent conversation activity and interactions among them, 
% such as \emph{posts}, \emph{comments}, \emph{mentions}, or \emph{endorsements}.
% For some of our analyses, 
% We also assume that the text regarding what each person has written about the topic is available.
% For the Twitter setting that we focus on in this work, we examine a few ways to build such a conversation graph. %it is clear that extracting the activity regarding a text query, or a hashtag, provides several ways to build a conversation graph, which we examine in Section~\ref{sec:graph_building}.
% Depending on the underlying social-media platform, 
% other ways of obtaining a conversation graph and defining a topic can be used 
% (e.g., mentions of a specific person on Twitter, references to a page or curated topic data 
% on Facebook,\footnote{\url{https://www.facebook.com/business/news/topic-data}}
% and thread titles on Reddit).

% Explain: how we do it
We represent a topic of discussion with a
\emph{conversation graph}.
In such a graph, vertices represent users, 
and edges represent conversation activity and interactions,
such as \emph{posts}, \emph{comments}, \emph{mentions}, or \emph{endorsements}.
Our working hypothesis is that
it is possible to analyze the conversation graph of a topic
to reveal how controversial the topic is.
In particular, we expect the conversation graph of a controversial topic 
to have a \emph{clustered structure}.
% Such clustering may be present at different levels, 
% for instance, in the graph structure formed by interactions among users,
% or according to the keywords used by different sides.
This hypothesis is based on the fact that a controversial topic entails 
different sides with opposing points of view, 
and individuals on the same side tend to endorse and amplify each other's arguments~\cite{adamic2005political, akoglu2014quantifying, conover2011political}.

Our main contribution is to test this hypothesis. 
We achieve this result by studying a large number of candidate features,  
based on the following \emph{aspects} of activity:
($i$) \emph{structure of endorsements}, i.e., who agrees with whom on the topic, 
($ii$) \emph{structure of the social network}, i.e., 
who is connected with whom among the participants in the conversation,
($iii$) \emph{content}, i.e., the keywords used in the topic,
($iv$) \emph{sentiment}, i.e., the tone (positive or negative) used to discuss the topic.
Our study shows that all features, except from content-based ones, 
are useful in detecting controversial topics, to different extents.
Particularly for Twitter, 
%  given also the ease of extraction, 
we find endorsement features (i.e., retweets) to be the most useful.

The extracted features are then used to compute the \emph{controversy score} of a topic.
We offer a systematic definition and provide a 
thorough evaluation of measures to quantify controversy.
We employ a broad range of topics, 
both controversial and non-controversial ones, 
on which we evaluate several measures, either defined in this paper or coming from the 
literature~\cite{guerra2013measure, morales2015measuring}.
We find that one of our newly-proposed measures, 
based on \emph{random walks}, 
is able to discriminate controversial topics with great accuracy.
In addition, it also generalizes well as it agrees with previously-defined measures 
when tested on datasets from existing work.
%while previous measures perform poorly on our testbed.
We also find that the \emph{variance} of the sentiment expressed on a topic is a reliable indication of controversy.
%In contrast, we find that the {\em modularity measure}~\cite{newman2006modularity}, a popular measure for identifying clustered structure in graphs, does not yield competitive results.

The approach to quantifying controversy %in social media
presented in this paper can be condensed into a three-stage pipeline:
($i$) building a \emph{conversation graph} among the users who contribute to a topic, 
where edges signify that two users are in some form of agreement,
($ii$) identifying the potential sides of the controversy from the graph structure or the textual content, and
($iii$) quantifying the amount of controversy in the graph.

% Outline
The rest of this paper is organized as follows.
Section~\ref{sec:related} discusses how this work fills gaps in the existing literature.
Subsequently, Section~\ref{sec:pipeline} provides a high level description of the pipeline for quantifying controversy of a topic, while Sections~\ref{sec:graph_building},~\ref{sec:partitioning}, and~\ref{sec:measure} detail each stage.
Section~\ref{sec:user-controversy} shows how to extend the controversy measures from topics to users who participate in the discussion.
We report the results of an extensive empirical evaluation of the proposed measures of controversy in Section~\ref{sec:experiments}.
Section~\ref{sec:content} extends the evaluation to a few measures that do not fit the pipeline.
We conclude in Section~\ref{sec:discussion} with a discussion on possible improvements and directions for future work, as well as lessons learned from carrying out this study.

%%%%%%%%%%%%%%%%%%%%%%%%%%%%%%%%%%%%%%%%%%%%%%%%%%%%%%%%%%%%%%%%%%

%%%%%%%%%%%%%%%%%%%%%%%%%%%%%%%%%%%%%%%%%%%%%%%%%%%%%%%%%%%%%%%%%%

\section{Related Work}
\label{sec:related}

%\enlargethispage{\baselineskip}

%%% Evidence for controversy in social networks and case studies
Analysis of controversy in online news and social media
has attracted considerable attention, 
and a number of papers have provided very interesting case studies. 
In one of the first papers, \citet{adamic2005political}
study the link patterns and discussion topics of political bloggers, 
focusing on blog posts about the U.S.\ presidential election of 2004.
They measure the degree of interaction between liberal and conservative blogs,
and provide evidence that  conservative blogs are linking to each other more frequently 
and in a denser pattern.
These findings %of \citeauthor{adamic2005political} 
are confirmed by the more recent study of \citet{conover2011political}, 
who also study controversy in political communication regarding congressional midterm elections.
Using data from Twitter, 
\citet{conover2011political} identify a highly segregated partisan structure 
(present in the retweet graph, but not in the mention graph), 
with limited connectivity between left- and right-leaning users.
In another recent work related to controversy analysis in political discussion,  
\citet{mejova2014controversy} identify a significant correlation between controversial issues 
and the use of negative affect and biased language. 

%%% methods to identify controversial topics

The papers mentioned so far study controversy in the political domain, 
and provide case studies centered around long-lasting major events,
such as presidential elections.
In this paper, we aim to identify and quantify controversy for any topic discussed in social media, 
% not only political issues, 
including short-lived and ad-hoc ones (for example, see the topics in Table~\ref{tab:datasets}).
The problem we study has been considered by previous work, 
but the methods proposed so far are, to a large degree, domain-specific.

The work of \citet{conover2011political}, 
discussed above, 
employs the concept of modularity and graph partitioning 
in order to verify (but not quantify) controversy structure of graphs extracted from
discussion of political issues on Twitter.
In a similar setting, \citet{guerra2013measure} propose an alternative graph-structure measure.
Their measure relies on the analysis of the boundary between two (potentially) polarized communities,
and performs better than modularity. 
%Our method is grounded in the same underlying principle as those previous works, 
%i.e., detecting cluster structure in graphs extracted from social interactions. 
Differently from these studies, our contribution consists in providing an extensive study of a large number of measures,
including the ones proposed earlier, 
and demonstrating clear improvement over those. 
We also aim at quantifying controversy 
in diverse and in-the-wild settings, 
rather than carefully-curated domain-specific datasets. 

In a recent study, \citet{morales2015measuring}
quantify polarity via the propagation of opinions of influential users on Twitter.
They validate their measure with a case study from Venezuelan politics. 
Again, our methods are not only more general and domain agnostic, 
but they provide more intuitive results. % than the method of \citeauthor{morales2015measuring}.
In a different approach, 
\citet{akoglu2014quantifying} proposes a polarization metric that uses signed bipartite opinion graphs.
The approach differs from ours as it relies on the availability of this particular type of data, 
which is not as readily available as social-interaction graphs.

\begin{table}[]
\centering
\caption{Summary of related work for identfying/quantifying controversial topics.% use tbl in the acmsmall package
\label{tab:related_quantifying}}
{\begin{tabular}{l cccc}
Paper & Identifying & Quantifying & Content & Network \\
\toprule
	\citet{choi2010identifying}      &    \cmark   &             &  \cmark &         \\
	\citet{popescu2010detecting}     &    \cmark   &             &  \cmark &         \\
	\citet{mejova2014controversy} &   \cmark    &             &   \cmark &         \\
	\citet{klenner2014verb}      &   \cmark    &             &   \cmark &         \\
	\citet{tsytsarau2011scalable}      &   \cmark    &             &    \cmark     &         \\
	\citet{dori2015automated}    &   \cmark    &             &   \cmark &         \\
	\citet{jang2016probabilistic} &   \cmark    &             &   \cmark &         \\
	\citet{conover2011political} & \cmark    &             &   &  \cmark      \\
	\citet{coletto2017motif} &   \cmark    &             &   &  \cmark      \\
	\citet{akoglu2014quantifying}     &   \cmark  &     &      &    \cmark     \\
	\citet{amin2017unveiling} & \cmark & & & \cmark \\
	\citet{guerra2013measure} & \cmark & \cmark & & \cmark \\
	\citet{morales2015measuring}     &  \cmark   &  \cmark           &      &    \cmark     \\
	\textbf{\citet{garimella2016quantifying}}      &   \cmark   &     \cmark        &        &  \cmark       \\
%      &             &             &         &        \\
\bottomrule
\end{tabular}}
\end{table}

Similarly to the papers discussed above, in our work 
we quantify controversy based on the graph structure of social interactions.
In particular, 
we assume that controversial and polarized topics induce graphs 
with clustered structure,
representing different opinions and points of view.
This assumption relies on the concept of ``echo chambers,'' 
which states that opinions or beliefs  stay inside communities created by like-minded people,
who reinforce and endorse the opinions of each other. 
This phenomenon has been quantified in many recent 
%studies~\cite{an2014partisan,bakshy2015exposure,flaxmanfilter,grevet2014managing}.%,hermida2011your}.
studies~\cite{an2014partisan,flaxmanfilter,grevet2014managing}.%,hermida2011your}.

A different direction for quantifying controversy, followed by~\citet{choi2010identifying}
and~\citet{mejova2014controversy},
%departing from the graph-structure methods discussed above,  
relies on text and sentiment analysis.
Both studies focus on language found in news articles. % reporting a topic.
In our case, since we are mainly working with Twitter, 
where text is short and noisy, 
and since we are aiming at quantifying controversy in a domain-agnostic manner, 
text analysis has its limitations. 
Nevertheless, we experiment with incorporating content features in our approach. 
% As we will see in a later section, 
%We find that our content-based analysis does not lead to good results, 
%while variance in the sentiment shows a good correlation with controversy, 
%and therefore can be used to enhance graph structure-based methods.

A summary of related work categorized along different dimensions is presented in Table~\ref{tab:related_quantifying}. As we mention above, most existing work to date tries to \emph{identify} controversial topics as case studies on a particular topic, either using content or networks of interactions. 
Our work is one of the few that \emph{quantifies} the degree of controversy using language and domain independent methods. 
Section~\ref{sec:experiments} shows that our method outperforms existing ones~\cite{guerra2013measure,morales2015measuring}. 

Finally, our findings on controversy 
have several potential applications on news-reading and public-debate scenarios.
Quantifying  controversy 
can provide a basis for analyzing the ``news diet'' of 
readers~\cite{kulshrestha2015characterizing,lacour2012balanced}, 
offering the chance of better information by 
providing recommendations of contrarian views~\cite{munson2013encouraging}, 
deliberating debates~\cite{esterling2010much},
and connecting people with opposing views~\cite{doris2013political,graells2013data}.

% Our measures could be potentially used to discover Twitter ``firestorms''~\cite{pfeffer2014understanding}.

%%%%%%%%%%%%%%%%%%%%%%%%%%%%%%%%%%%%%%%%%%%%%%%%%%%%%%%%%%%%%%%%%%

%%%%%%%%%%%%%%%%%%%%%%%%%%%%%%%%%%%%%%%%%%%%%%%%%%%%%%%%%%%%%%%%%%

\section{Pipeline}
\label{sec:pipeline}

\begin{figure}[b]
\begin{center}
\includegraphics[width=\columnwidth]{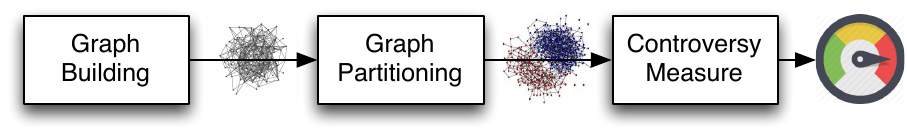}
\caption{Block diagram of the pipeline for computing controversy scores.}
\label{fig:block}
\end{center}
\vspace{-\baselineskip}
\end{figure}

% introduce general concepts of topic, partition, controversy, and the kind of output we want.
Our approach to measuring controversy is based on a systematic way of 
characterizing social media activity. We employ 
a pipeline with three stages, namely \textit{graph building}, \textit{graph 
partitioning}, and \textit{controversy measure}, as shown in Figure~\ref{fig:block}. 
The final output of the pipeline is a value
% between zero and one 
that measures how controversial a topic is, with higher values 
corresponding to higher degree of controversy. We provide a high-level 
description of each stage here and more details in the sections 
that follow.

%\note[gdfm]{This paragraph should probably go further down in the graph building, also could be more formal.}
%}
%\note[michael]{Does ``at least two retweets per hashtag" mean that we use the average value per hashtag? If so, the claim about the ``union of the retweet graphs" does not hold.}

\subsection{Building the Graph}
\label{sec:pipeline_graph}
% topic
The purpose of this stage is to build a \emph{conversation graph} that
represents activity related to a single \emph{topic} of discussion. 
% A topic generally captures the discussion around a single news event.
In our pipeline, a topic is operationalized as a set of related hashtags (details in \S\ref{sec:hashtags_to_clusters}),
and the social media activity related to the topic consists of those items (e.g., posts) that match this set of hashtags.
For example, the query might consist simply of a keyword,
such as ``\#ukraine", in which case the related activity consists of all tweets that contain that keyword,
or related tags such as \#kyiv and \#stoprussianaggression.
%\todo{Insert two examples of hashtags related to \#ukraine found by our method}
Even though we describe textual queries in standard document-retrieval form, in principle queries can take other forms, as long as they are able to induce a graph from the social media activity (e.g., RDF queries, or topic models).

% Explain: how we do it
Each item related to a topic is associated with one user who generated it, and we build a graph where each user who contributed to the topic is assigned to one vertex.
In this graph, an edge between two vertices represents \emph{endorsment}, \emph{agreement}, or \emph{shared point of view} between the corresponding users.
Section~\ref{sec:graph_building} details several ways to build such a graph. %, taking into account either interactions between users, the social connections between them, or the similarity of the messages they generate.

% partition
\subsection{Partitioning the Graph}
\label{sec:pipeline_partition}
In the second stage, the resulting conversation graph 
is fed into a \emph{graph partitioning} algorithm to extract \emph{two} partitions (we defer considering multi-sided controversies to further study). Intuitively, the % partitioning is performed in such a way, so that each of the two
two partitions correspond to two disjoint sets of users who possibly belong to different sides in the discussion. % -- assuming that there actually exactly two sides in this discussion.
In other words, the output of this stage answers the following question:
%``considering which users share the same point of view on this topic, and 
``assuming that users are split into two sides according to their point of view on the topic, 
which are these two sides?''
Section~\ref{sec:partitioning} describes this stage in further detail.
%The rationale for obtaining such a partition is that,
If indeed there are two sides which do not agree with each other --a controversy--
then the two partitions should be loosely connected to each other, given the semantic of the edges.
This property is captured by a measure computed in the third and final stage of the pipeline.

% controversy
\subsection{Measuring Controversy}
The third and last stage takes as input the graph built by the first stage and partitioned by the second stage, and computes the value of a \emph{controversy measure} %as one feature 
that characterizes how controversial the topic is.
Intuitively, a controversy measure aims to capture how sell separated the two partitions are.
We test several such measures, including ones based on random walks, betweenness centrality, and low-dimensional embeddings.
Details are provided in Section~\ref{sec:measure}.

%\mpara{Notes:} {\bf 1.} We stress that the final value generated by the pipeline
%depends on the algorithms of choice employed at each of the three stages.
% Therefore, the total number of features generated by this pipeline for one topic
% is equal to the number of combinations of algorithms across the three stages.
%{\bf 2.} We've made the implicit assumption that for each
%topic there are either one or two `sides' arguing for each topic -- in the latter
%case, the topic is deemed controversial.
%This assumption matches well the datasets that we've worked with, but obviously
%it need not be the case generally, as there may be more than one sides for
%one controversial topic. We come back to this issue in
%Section~\ref{sec:generalizability} and discuss how such cases can be
%accomodated in future work.

%%%%%%%%%%%%%%%%%%%%%%%%%%%%%%%%%%%%%%%%%%%%%%%%%%%%%%%%%%%%%%%%%%

%%%%%%%%%%%%%%%%%%%%%%%%%%%%%%%%%%%%%%%%%%%%%%%%%%%%%%%%%%%%%%%%%%

\section{Graph Building}
\label{sec:graph_building}

\begin{table*}[t]
\caption{Datasets statistics: hashtag, sizes of the follow and retweet graphs, and description of the event. The top group represent controversial topics, while the bottom one represent non-controversial ones.}
\label{tab:datasets}
\centering
\small
\resizebox{\linewidth}{!}{%
\begin{tabular}{l r r r r r l}
\toprule
Hashtag & \# Tweets & \multicolumn{2}{c}{Retweet graph} & \multicolumn{2}{c}{Follow graph} & Description and collection period (2015)\\
\cmidrule(lr){3-4} \cmidrule(lr){5-6}
 & & $|V|$ & $|E|$ & $|V|$ & $|E|$ & \\
\midrule
\#beefban & \num{422908} & \num{21590} & \num{30180} & \num{9525} & \num{204332} & Government of India bans beef, Mar 2--5\\
\#nemtsov & \num{371732} & \num{43114} & \num{77330} & \num{17717} & \num{155904} & Death of Boris Nemtsov, Feb 28--Mar 2\\
\#netanyahuspeech & \num{1196215} & \num{122884} & \num{280375} & \num{49081} & \num{2009277} & Netanyahu's speech at U.S. Congress, Mar 3--5\\
\#russia\_march & \num{317885} & \num{10883} & \num{17662} & \num{4844} & \num{42553} & Protests after death of Boris Nemtsov (``march''), Mar 1--2\\
\#indiasdaughter & \num{776109} & \num{68608} & \num{144935} & \num{38302} & \num{131566} & Controversial Indian documentary, Mar 1--5\\
\#baltimoreriots & \num{1989360} & \num{289483} & \num{432621} & \num{214552} & \num{690944} & Riots in Baltimore after police kills a black man, Apr 28--30\\
\#indiana & \num{972585} & \num{43252} & \num{74214} & \num{21909} & \num{880814} & Indiana pizzeria refuses to cater gay wedding, Apr 2--5\\
\#ukraine & \num{514074} & \num{50191} & \num{91764} & \num{31225} & \num{286603} & Ukraine conflict, Feb 27--Mar 2\\
\#gunsense & \num{1022541} & \num{30096} & \num{58514} & \num{17335} & \num{841466} & Gun violence in U.S., Jun 1--30\\
\#leadersdebate & \num{2099478} & \num{54102} & \num{136290} & \num{22498} & \num{1211956} & Debate during the U.K. national elections, May 3\\
\midrule
\#sxsw & \num{343652} & \num{9304} & \num{11003} & \num{4558} & \num{91356} & SXSW conference, Mar 13--22\\
\#1dfamheretostay & \num{501960} & \num{15292} & \num{26819} & \num{3151} & \num{20275} & Last OneDirection concert, Mar 27--29\\
\#germanwings & \num{907510} & \num{29763} & \num{39075} & \num{2111} & \num{7329} & Germanwings flight crash, Mar 24--26\\
\#mothersday & \num{1798018} & \num{155599} & \num{176915} & \num{2225} & \num{14160} & Mother's day, May 8\\
\#nepal & \num{1297995} & \num{40579} & \num{57544} & \num{4242} & \num{42833} & Nepal earthquake, Apr 26--29\\
\#ultralive & \num{364236} & \num{9261} & \num{15544} & \num{2113} & \num{16070} & Ultra Music Festival, Mar 18--20\\
\#FF & \num{408326} & \num{5401} & \num{7646} & \num{3899} & \num{63672} & Follow Friday, Jun 19\\
\#jurassicworld & \num{724782} & \num{26407} & \num{32515} & \num{4395} & \num{31802} & Jurassic World movie, Jun 12-15\\
\#wcw & \num{156243} & \num{10674} & \num{11809} & \num{3264} & \num{23414} & Women crush Wednesdays, Jun 17\\
\#nationalkissingday & \num{165172} & \num{4638} & \num{4816} & \num{790} & \num{5927} & National kissing day, Jun 19\\\bottomrule
\end{tabular}}
%\vspace{-\baselineskip}
\end{table*}

This section provides details about different approaches to build graphs from raw data. % (first stage of pipeline -- Section~\ref{sec:graph_details}).
%\subsection{Graph building}
%\label{sec:graph_details}
We use posts on Twitter to create our datasets.\footnote{From the full Twitter firehose stream.}
Twitter is a natural choice for the problem at hand, as it represents one of the main fora for public debate in online social media, and is often used to report news about current events.
Following the procedure described in Section~\ref{sec:pipeline_graph},
we specify a set of queries (indicating topics), and build one graph for each query.
We choose a set of topics which is balanced between controversial and non-controversial ones, so as to test for both false positives and false negatives.

We use Twitter hashtags as \emph{queries}.
% Hashtags are a special type of keyword that begins with the hash character (`\#' ~-- e.g., `\#beefban', `\#ukraine', `\#mothersday' ).
Users commonly employ hashtags to indicate the topic of discussion their posts pertain to.
Then, we define a \emph{topic} as the set of hashtags related to the given query.
Among the large number of hashtags that appear in the Twitter stream,
we consider those that were trending during the period from Feb 27 to Jun 15, 2015.
By manual inspection we find that most trending hashtags are not related to controversial discussions~\cite{garimella2016exploring}.

We first manually pick a set of $10$ hashtags that we know
represent \emph{controversial} topics of discussion. All hashtags
in this set have been widely covered by mainstream media, and
have generated ample discussion, both online and offline.
Moreover, to have a dataset that is balanced between
controversial and non-controversial topics,
we sample another set of $10$ hashtags that %we know
represent \emph{non-controversial} topics of discussion.
These hashtags are related mostly to ``soft news'' and
entertainment, but also to events that, while being impactful and 
dramatic, did not generate large controversies (e.g., \#nepal and 
\#germanwings).
In addition to our intuition that these topics are non-controversial, we manually check a sample of tweets, and we are unable to identify any clear instance of controversy.\footnote{Code and networks used in this work are available at \url{http://github.com/gvrkiran/controversy-detection}.}

As a first step, we now describe the process of expanding a single hashtag into a set of related hashtags which define the topic.
The goal of this process is to broaden the definition of a topic,
and ultimately improve the coverage of the topic itself.

\subsection{From hashtags to topics}
\label{sec:hashtags_to_clusters}
In the literature, a topic is often defined by a single hashtag.
However, this choice might be too restrictive in many cases. 
For instance, the opposing sides of a controversy might use different hashtags, as the hashtag itself is loaded with meaning and used as a means to express their opinion.
Using a single hashtag may thus miss part of the relevant posts.

To address this limitation, we extend the definition of topic to be more encompassing.
Given a \emph{seed} hashtag, we define a topic as a set of related hashtags, which co-occur with the seed hashtag.
To find related hashtags, we employ (and improve upon) a recent clustering algorithm tailored for the purpose~\cite{feng2015streamcube}.

\citet{feng2015streamcube} develop a simple measure to compute the similarity between two hashtags, which relies on co-occurring words and hashtags.
The authors then use this similarity measure to find closely related hashtags and define clusters.
However, this simple approach presents one drawback, in that very popular hashtags such as \#ff or \#follow co-occur with a large number of hashtags.
Hence, directly applying the original approach results in extremely noisy clusters.
Since the quality of the topic affects critically the entire pipeline, we want to avert this issue and ensure minimal noise is introduced in the expanded set of hashtags.

Therefore, we improve the basic approach by taking into account and normalizing for the popularity of the hashtags.
Specifically, we compute the document frequency of all hashtags on a random 1\% sample of the Twitter stream,\footnote{From the Twitter Streaming API \url{https://dev.twitter.com/streaming/reference/get/statuses/sample}.} and normalize the original similarity score between two hashtags by the inverse document frequency.
The similarity score is formally defined as
\begin{equation}
\label{eq:hashtag_similarity}
\similar(h_s,h_t) = \frac{1}{1+\log(df(h_t))} 
\left( \alpha \, \cos(W_s,W_t) + (1-\alpha) \, 
\cos(H_s, H_t) \right) ,
\end{equation}
where $h_s$ is the seed tag, $h_t$ is the candidate tag, $W_x$ and $H_x$ are the sets of words and hashtags that co-occur with hashtag $h_x$, respectively, $\cos$ is the cosine similarity between two vectors, 
$df$ is the document frequency of a tag,
and $\alpha$ is a parameter that balances the importance of words compared to hashtags in a post.
% (we used $\alpha$ = 0.3, as was proposed in~\cite{feng2015streamcube}).
%\note[gdfm]{If alpha + beta = 1, then we should say alpha and (1-alpha) and have a single parameter. Also, report used (default?) values for the parameters. Also what is the value of $k$?}
	
By using the similarity function in Equation~\ref{eq:hashtag_similarity}, we retrieve the top-$k$ most similar hashtags to a given seed.
The set of these hashtags along with the initial seed defines the topic for the given seed hashtag.
%(e.g. for \#netanyahuspeech, the other hashtags we got after the clustering were \#istandwithisrael, \#bibispeech, \#iran, \#bibispeaks4me, etc).
The topic is used as a filter to get all tweets which contain at least one of the hashtags in the topic.
In our experiments we use $\alpha$ = 0.3 (as proposed by~\citet{feng2015streamcube}) and $k$ = 20.

\begin{figure}[t]
\begin{minipage}{.49\linewidth}
\centering
\subfloat[]{\label{}\includegraphics[width=\textwidth]{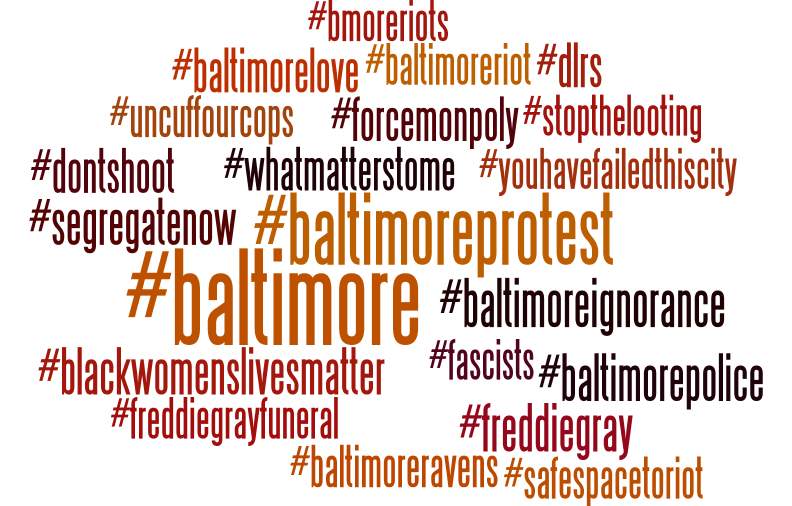}}
\end{minipage}%
\hspace{0.01\textwidth}
\begin{minipage}{.49\linewidth}
\centering
\subfloat[]{\label{}\includegraphics[width=\textwidth]{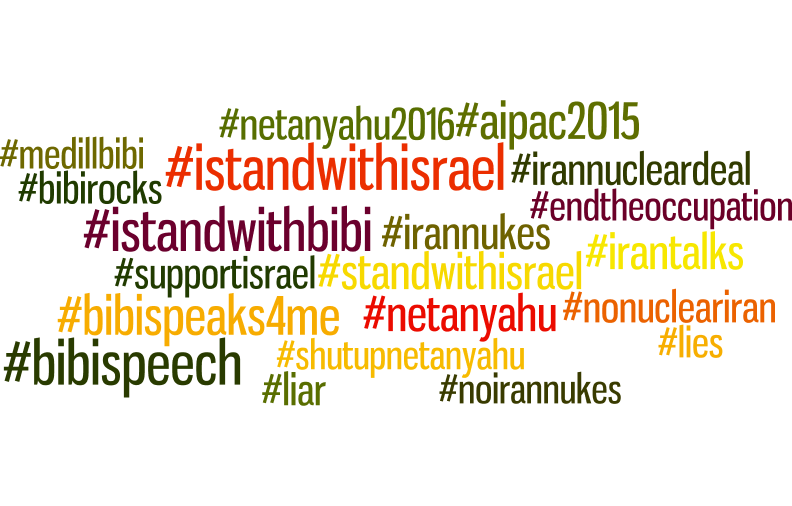}}
\end{minipage}\par\medskip
\caption{Sets of related hashtags for the topics (a) \#baltimoreriots and (b) \#netanyahuspeech.}
\label{fig:similar_hashtags}
%\vspace{-\baselineskip}
\end{figure}

Figure~\ref{fig:similar_hashtags} shows the top-20 most similar hashtags for two different seeds: (a) \#baltimoreriots, which identifies the discussion around the Baltimore riots against police violence in April 2015 and (b) \#netanyahuspeech, which identifies the discussion around Netanyahu's speech at the US congress  in March 2015.
By inspecting the sets of hashtags, it is possible to infer the nature of the controversy for the given topic, as both sides are represented.
For instance, the hashtags \#istandwithisrael and \#shutupbibi represent opposing sides in the dicussion raised by Netanyahu's speech.
Both hashtags are recovered by our approach when \#netanyahuspeech is provided as the seed hashtag.
It is also clear why using a single hashtag is not sufficient to define a topic: the same user is not likely to use both \#safespacetoriot and \#segregatenow, even though the two hashtags refer to the same event (\#baltimoreriots).

% -- Done section on expanding hashtag sets

\subsection{Data aspects}
For each topic, we retrieve all tweets that contain one of its hashtags and that are
generated during the observation window.
%We only obtained additional hashtags (as described above) for controversial hashtags.
%Each hashtag along with its set of related tweets define a single dataset.
We also ensure that the selected hashtags are associated with a 
large enough volume of activity.
Table~\ref{tab:datasets} presents the final set of seed hashtags,
along with their description and the number of related tweets.\footnote{We use a hashtag in Russian, \begin{minipage}{3.35em}\includegraphics{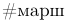}\end{minipage} ~,~which we refer to as \#russia\_march henceforth, for convenience.}
For each topic, we build a graph $G$ where we assign a vertex to each user who contributes to it,
%To build a graph $G$ for each hashtag, we assign one vertex to each user who employs the hashtag, 
and generate edges according to one of the following four approaches,
which capture different \emph{aspects} of the data source.

\spara{1. Retweet graph.}
%In Twitter, users are allowed to retweet content generated by another user.
Retweets typically indicate endorsement.\footnote{We do not consider `quote retweets' (retweet with a comment added) in our analysis.}
Users who retweet signal endorsement of the opinion expressed in the original tweet by propagating it further.
Retweets are not constrained to occur only between
users who are connected in Twitter's social network, but users are
allowed to retweet posts generated by any other user.

We select the edges for graph $G$ based on the retweet activity in the topic:
an edge exists between two users $u$ and $v$ if there are at least \emph{two} ($\tau = 2$) retweets between them that use the hashtag, irrespective of direction. % (i.e., whether $u$ retweeted $v$ or vice-versa).
We remark that, in preliminary experimentation with this approach, building the retweet graph with a threshold $\tau = 1$ did not produce reliable results.
We presume that a single retweet on a topic is not enough of a signal to infer endorsement.
Using $\tau = 2$ retweets as threshold proves to be a good trade-off between high selectivity (which hinders analysis) and noise reduction.
The resulting size for each retweet graph is listed in Table~\ref{tab:datasets}.

In an earlier version of this work~\cite{garimella2016quantifying}, when building a conversation graph for a single hashtag, we created an edge between two vertices only if there were ``at least two retweets per edge'' (in either direction) between the corresponding pair of users.
When defining topics as sets of hashtags, there are several ways to generalize this filtering step.
The simplest approach considers ``two of any'' in the set of hashtags that defines the topic.
However, this approach is too permissive, and results in an overly-inclusive graph, with spurious relationships and a high level of noise.
Instead, we opt to create an edge between two nodes only if there are at least two retweets for any given hashtag between the corresponding pair of users. 
In other words, the resulting conversation graph for the topic is the union of the retweet graphs for each hashtag in the topic, considered (and filtered) separately.

\spara{2. Follow graph.}
%In Twitter, a user $u$ can create social connections to other users $v$ and thus receive automatically the posts they create --
%in the platform's lingo, a user $u$ %that creates a social connection to another user $v$ 
%is said to ``follow'' $v$.
In this approach, we build the follow graph induced by a given hashtag.
We select the edges for graph $G$ based on the social connections between Twitter users \emph{who employ the given hashtag}:
an edge exists between users $u$ and $v$ if $u$ follows $v$ or vice-versa.
We stress that the graph $G$ built with this approach is topic-specific,
as the edges in $G$ are constrained to connections
between users who discuss the topic that is specified as input
to the pipeline.

The rationale for using this graph is based on an assumption of the
presence of homophily in the social network, which is a common trait in
this setting.
To be more precise, we expect that \emph{on a given topic} people will agree more often than not with
people they follow, and that for a controversial topic
this phenomenon will be reflected in well-separated partitions of the
resulting graph.
Note that using the entire social graph would not necessarily produce 
well-separated partitions that correspond to single topics of discussion, as those
partitions would be ``blurred'' by the existence of additional edges that
are due to other reasons (e.g., offline social connections).
% Note that this is not necessarily the case for the entire social graph,
% but it is more likely when restricted to people engaged in a discussion on the same topic.

On the practical side, while the retweet information is readily available in the stream of tweets, the social network of Twitter is not.
Collecting the follower graph thus requires an expensive crawling phase.
The resulting graph size for each follow graph is listed in Table~\ref{tab:datasets}.

\spara{3. Content graph.}
We create the edges of graph $G$ based on whether
users post instances of the same content. 
Specifically, we experiment with the following three variants:
create an edge between two vertices if the users
($i$) use the same hashtag, other than the ones that defines the topic,
($ii$) share a link to the same URL, or ($iii$) share a link with the same
URL domain (e.g., \url{cnn.com} is the domain for all pages on the website of CNN).

\spara{4. Hybrid content \& retweet graph.}
We create edges for graph $G$ according to a state-of-the-art process that blends content and graph information~\cite{ruan2013efficient}.
Concretely, we associate each user with a vector of frequencies of mentions for different hashtags.
Subsequently, we create edges between pairs of users whose corresponding vectors have high cosine similarity, and combine them with edges from the retweet graph, built as described above.
For details, we refer the interested reader to the original publication~\cite{ruan2013efficient}.

% Given that by homophily people tend to follow people that are similar to them, we can expect that, on a given topic, people tend to agree with people they follow.
% Note that this is not necessarily the case for the whole social graph, but it is more likely when restricted to people engaged in a discussion on a topic.
% This kind of topic-induced social graph has structural properties that are quite different from the retweet graph, as it is generated by a different process.
% Nevertheless, we show that also this kind of graph is a useful input to detect controversial topics.

\begin{figure*}[t]
\begin{minipage}{.24\linewidth}
\centering
\subfloat[]{\label{}\includegraphics[width=\textwidth, height=\textwidth]{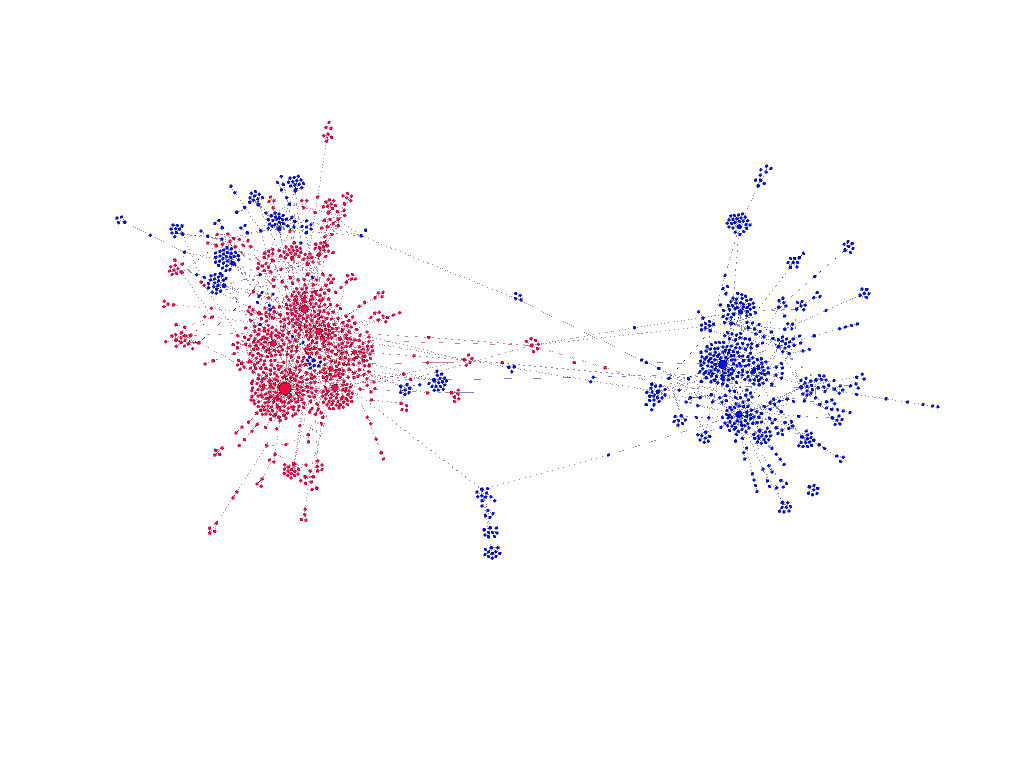}}
\end{minipage}%
\begin{minipage}{.24\linewidth}
\centering
\subfloat[]{\label{}\includegraphics[width=\textwidth, height=\textwidth]{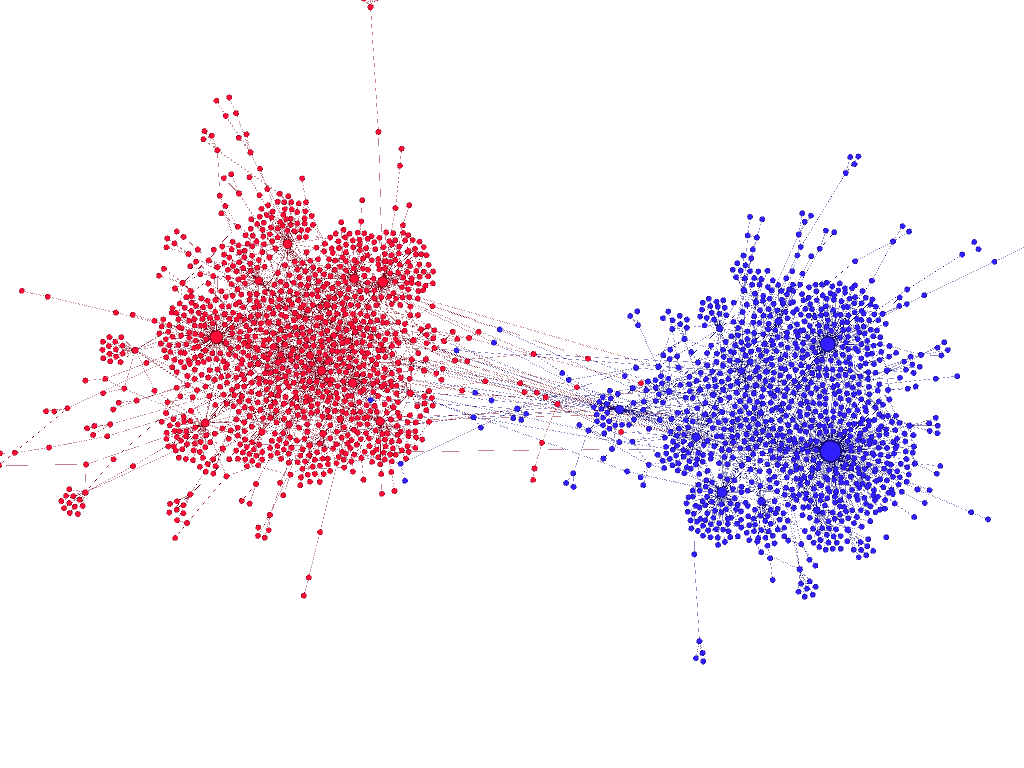}}
\end{minipage}
\begin{minipage}{.24\linewidth}
\centering
\subfloat[]{\label{}\includegraphics[width=\textwidth, height=\textwidth]{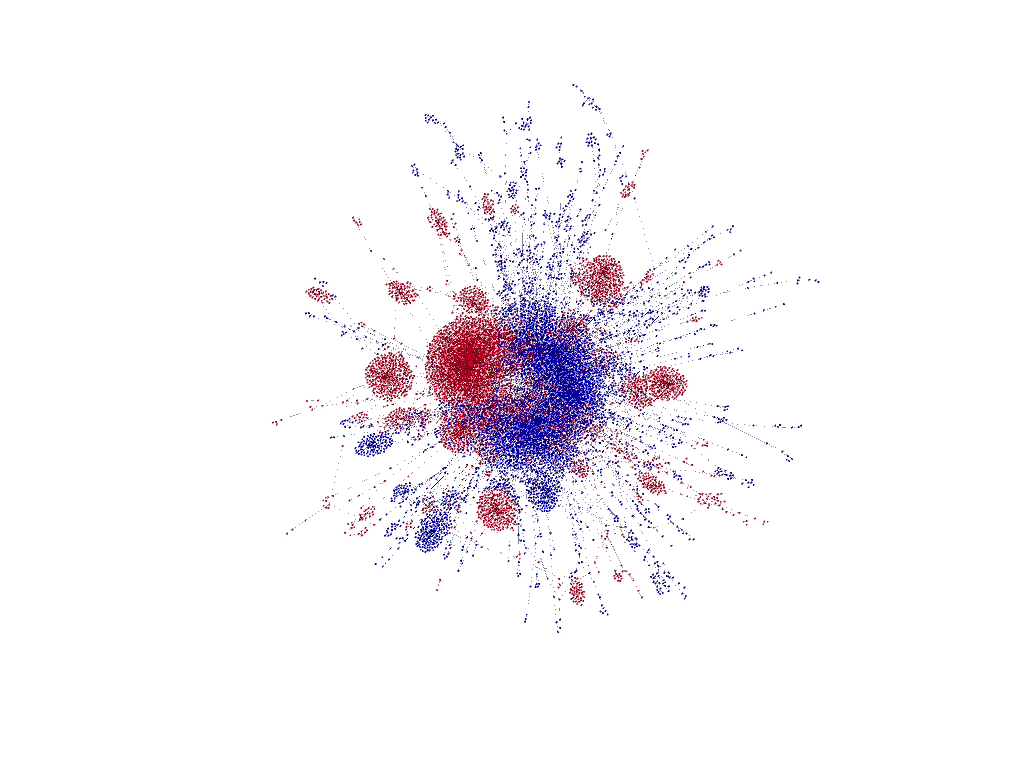}}
\end{minipage}
\begin{minipage}{.24\linewidth}
\centering
\subfloat[]{\label{}\includegraphics[width=\textwidth, height=\textwidth]{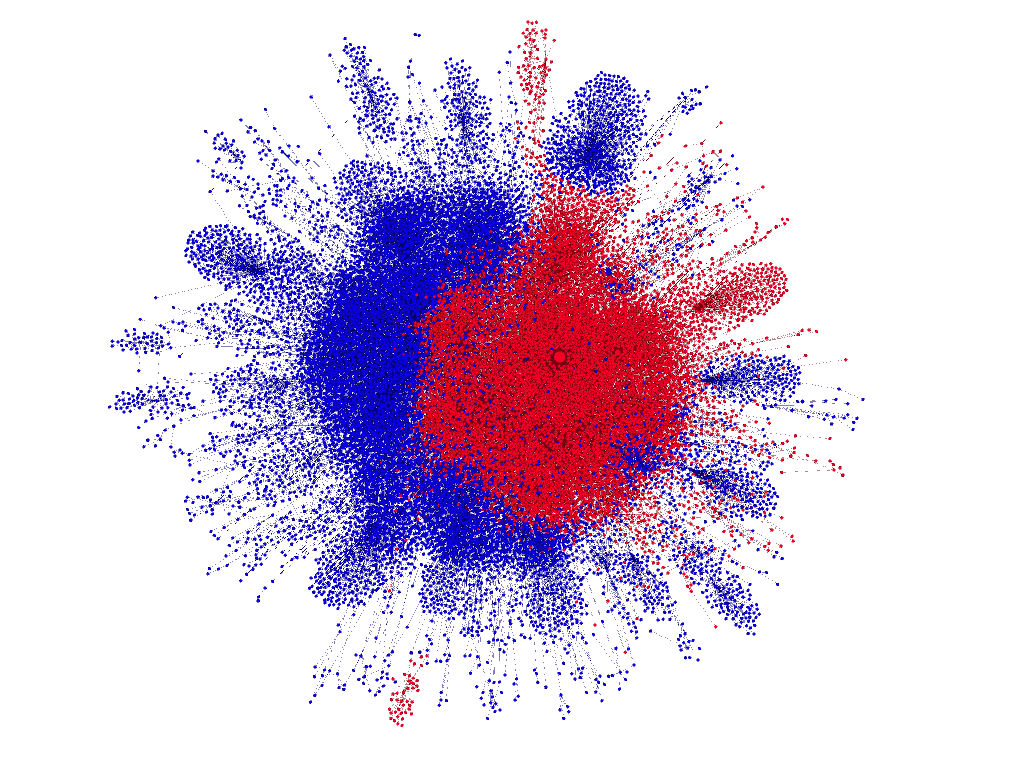}}
\end{minipage}\par\medskip
\begin{minipage}{.24\linewidth}
\centering
\subfloat[]{\label{}\includegraphics[width=\textwidth, height=\textwidth]{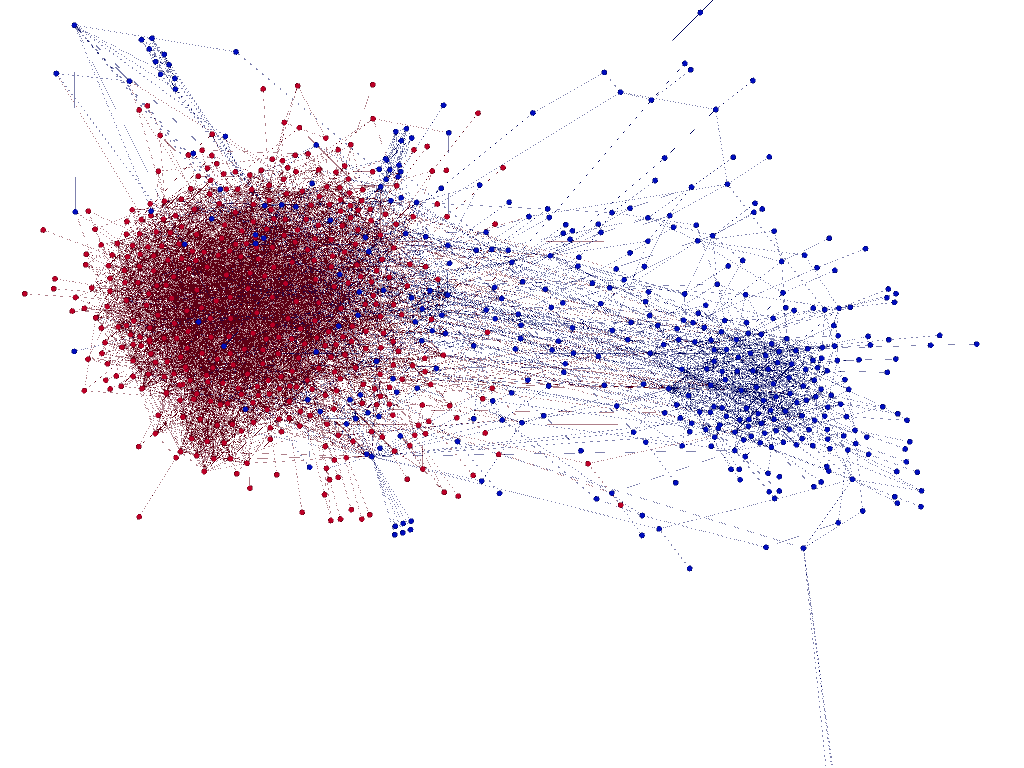}}
\end{minipage}
\begin{minipage}{.24\linewidth}
\centering
\subfloat[]{\label{}\includegraphics[width=\textwidth, height=\textwidth]{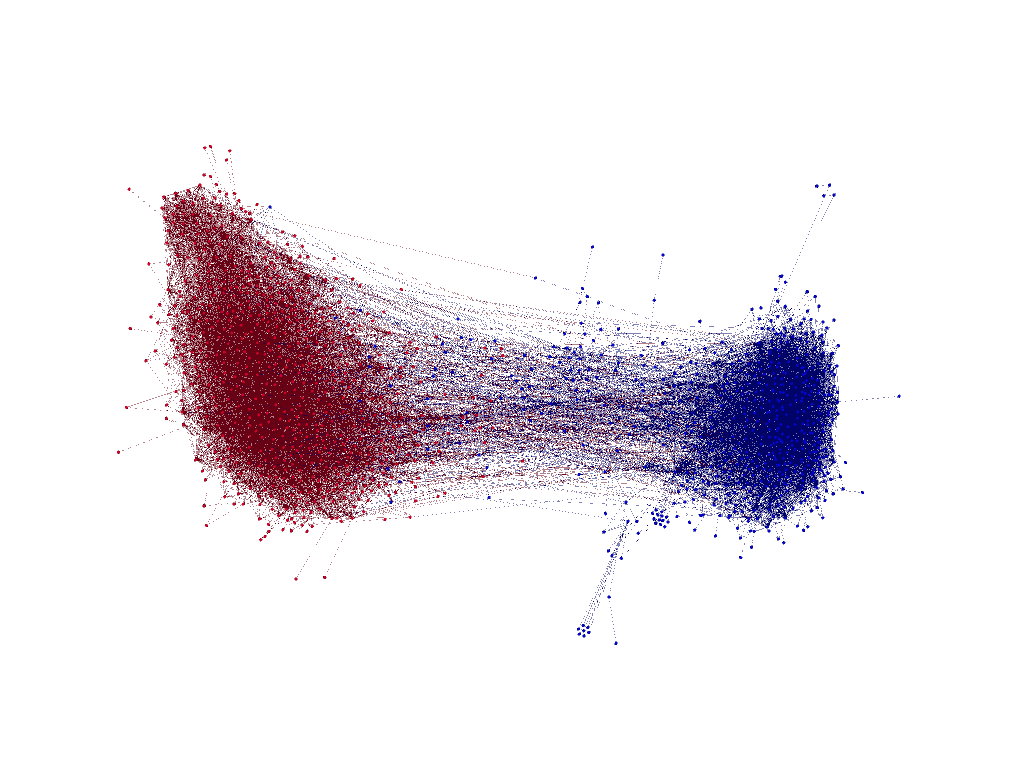}}
\end{minipage}
\begin{minipage}{.24\linewidth}
\centering
\subfloat[]{\label{}\includegraphics[width=\textwidth, height=\textwidth]{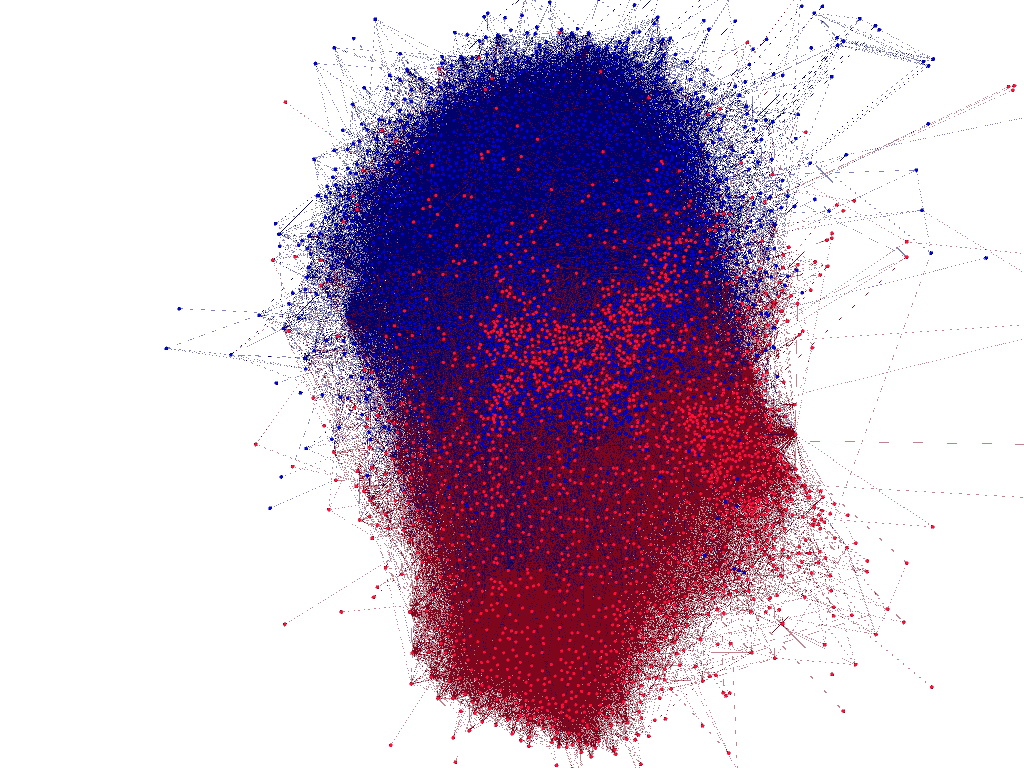}}
\end{minipage}
\begin{minipage}{.24\linewidth}
\centering
\subfloat[]{\label{}\includegraphics[width=\textwidth, height=\textwidth]{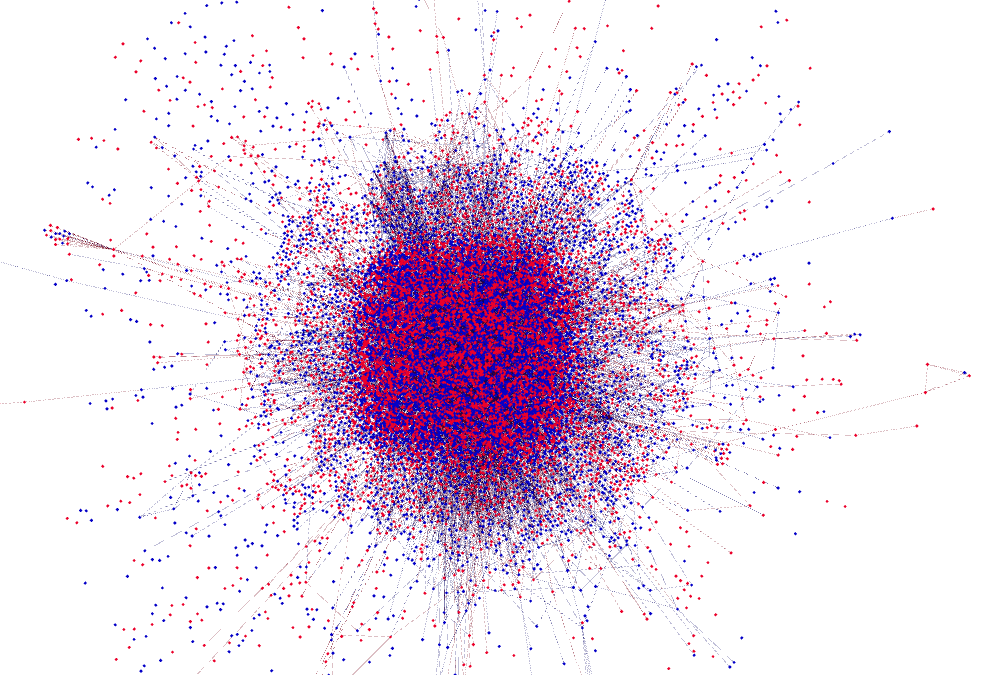}}
\end{minipage}\par\medskip
\caption{Sample conversation graphs with retweet (top) and follow (bottom) aspects (visualized using the force-directed layout algorithm in Gephi). The left side is controversial, (a,e) \#beefban, (b,f) \#russia\_march, while the right side is non-controversial, (c,g) \#sxsw, (d,h) \#germanwings. Only the largest connected component is shown.}
\label{fig:visualization}
%\vspace{-\baselineskip}
\end{figure*}

\smallskip
\section{Graph partitioning}
\label{sec:partitioning}

%This section shows examples of graph partitions obtained via METIS. % (second stage of pipeline -- Section~\ref{sec:partitioning})
As previously explained, 
% in Section~\ref{sec:pipeline_partition}, 
we use a graph partitioning algorithm 
%in the second stage of the pipeline 
to produce two partitions on the conversation graph.
% built in the first stage.
To do so, we rely on a state-of-the-art off-the-shelf algorithm,
% for this task, 
METIS~\cite{karypis1995metis}.
Figure~\ref{fig:visualization} displays the two partitions returned for some of the topics on their corresponding retweet and follow graphs (Figures~\ref{fig:visualization}(a)-(d) and Figures~\ref{fig:visualization}(e)-(h), respectively).\footnote{Other topics show similar trends.}
The partitions are depicted in blue or red.
The graph layout is produced by Gephi's ForceAtlas2 algorithm~\cite{jacomy2014forceatlas2}, and is based solely on the structure of the graph, not on the partitioning computed by METIS. Only the largest connected component is shown in the visualization, though in all the cases the largest connected component contains more than $90\%$ of nodes.

From an initial visual inspection of the partitions identified on retweet and follow graphs, we find that the partitions match well with our intuition of which topics are controversial (the partitions returned by METIS are well separated for controversial topics).
To make sure that this initial assessment of the partitions is not an artifact of the visualization algorithm we use, we try other layouts offered by Gephi.
In all cases we observe similar patterns.
We also manually sample and check tweets from the partitions, to verify the presence of controversy.
While this anecdotal evidence is hard to report, indeed the partitions seem to capture the spirit of the controversy.\footnote{For instance, of these two tweets for \#netanyahuspeech from two users on opposing sides, one is clearly supporting the speech \url{https://t.co/OVeWB4XqIg}, while the other highlights the negative reactions to it \url{https://t.co/v9RdPudrrC}.}

\begin{figure}[t]
\begin{minipage}{.49\linewidth}
\centering
\subfloat[]{\label{}\includegraphics[width=\textwidth, height=\textwidth]{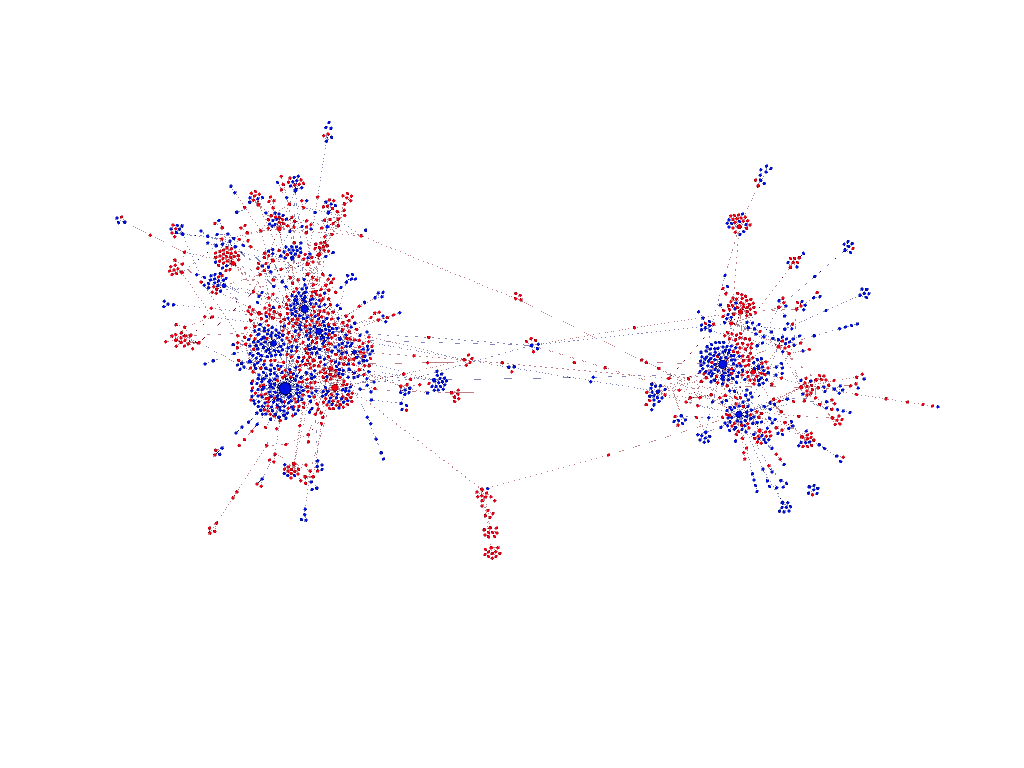}}
\end{minipage}%
\hspace{0.01\textwidth}
\begin{minipage}{.49\linewidth}
\centering
\subfloat[]{\label{}\includegraphics[width=\textwidth, height=\textwidth]{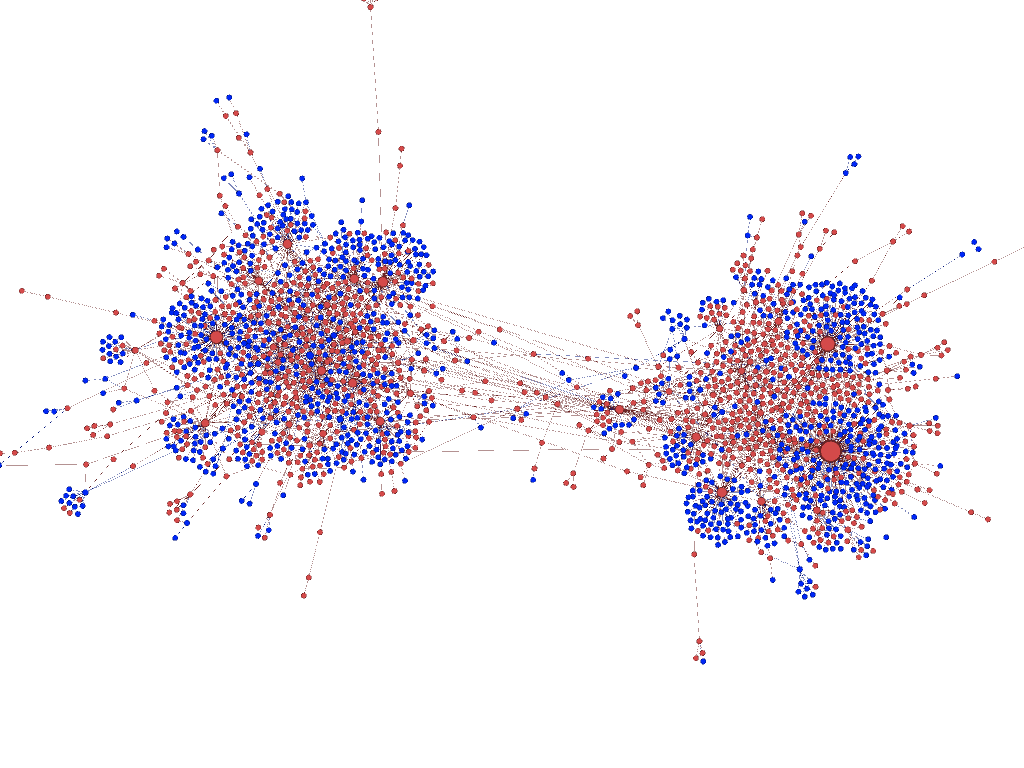}}
\end{minipage}\par\medskip
\caption{Partitions obtained for (a) \#beefban, (b) \#russia\_march by using the hybrid graph building approach. The partitions are more noisy than those in Figures~\ref{fig:visualization}(a,b).}
\label{fig:codicil}
%\vspace{-\baselineskip}
\end{figure}

On the contrary, the partitions identified on content graphs fail to match our intuition. %, as shown by their visual layout.
All three variants of the content-based approach lead to sparse graphs and highly overlapping partitions, even in cases of highly controversial issues.
The same pattern applies for the hybrid approach, as shown in Figure~\ref{fig:codicil}.
We also try a variant of the hybrid graph approach with vectors that represent the frequency of different URL domains mentioned by a user, with no better results.
We thus do not consider these approaches to graph building any further in the remainder of this paper.

Finally, we try graph partitioning algorithms of other types.
 % of different classes.
Besides METIS (cut based), we test spectral clustering, label propagation, and affiliation-graph-based models.
The difference among these methods is not significant, however from visual inspection METIS generates the cleanest partitions.

%%%%%%%%%%%%%%%%%%%%%%%%%%%%%%%%%%%%%%%%%%%%%%%%%%%%%%%%%%%%%%%%%%

%%%%%%%%%%%%%%%%%%%%%%%%%%%%%%%%%%%%%%%%%%%%%%%%%%%%%%%%%%%%%%%%%%

\section{Controversy Measures}
\label{sec:measure}
\enlargethispage{0.5\baselineskip}

This section describes the controversy measures used in this work.
For completeness, we describe both those measures proposed by us (\S\ref{sec:random_walk}, \ref{sec:betweenness}, \ref{sec:embedding}) as well as the ones from the literature that we use as baselines (\S\ref{sec:boundary}, \ref{sec:dipole}).

\subsection{Random walk}
\label{sec:random_walk}
This measure uses the notion of random walks on graphs.
It is based on the rationale that, in a controversial discussion, there are authoritative users on both sides, as evidenced by a large degree in the graph. 
The measure captures the intuition of how likely a random user on either side is to be exposed to authoritative content from the opposing side.

Let $G(V,E)$ be the graph built by the first stage and its two partitions $X$ and $Y$, ($X\cup Y = V$, $X\cap Y=\emptyset$) identified by the second stage of the pipeline.
We first distinguish the $k$ \emph{highest-degree vertices} from each partition.
High degree is a proxy for authoritativeness, as it means that a user has received a large number of endorsements on the specific topic.
Subsequently, we select one partition at random (each with probability $0.5$)
and consider a random walk that starts from a random vertex in that
partition. The walk terminates when it visits any high-degree vertex (from either side).
% We repeat a large number of such walks ($1000$ for the results we report), starting an {\it equal} number of times from each partition.

\enlargethispage{0.5\baselineskip}
We define the \texttt{Random Walk Controversy} 
(\rwc) measure as follows. 
\emph{``Consider two random walks, one ending in partition $X$ and one ending in partition
$Y$, \rwc is the difference of the probabilities of
two events: (i) both random walks started from the partition they ended in
and (ii) both random walks started in a partition other than the one they
ended in.''}
The measure is quantified as
\begin{equation}
	\rwc = P_{_{XX}}P_{_{YY}} - P_{_{YX}}P_{_{XY}},
\end{equation}
where $P_{AB}$, $A,B\in \{X,Y\}$ is the conditional probability
\begin{equation}
	P_{AB} = \prob[\text{start in partition } A \mid \text{end in partition } B].
\end{equation}
%Note that 
The aforementioned probabilities have the
following desirable properties: ($i$) they are not
skewed by the size of each partition, as the random walk starts with equal probability from each partition, and
($ii$) they are not skewed by the total
degree of vertices in each partition, as % the starting vertex for a given partition is picked uniformly at random.
the probabilities are conditional on ending in either partition (i.e., the fraction of random walks ending in each partition is irrelevant).
\rwc is close to one when the probability of crossing sides is low, and close to zero when the probability of crossing sides is comparable to that of staying on the same side.

\subsection{An efficient variant of the random walk controversy score}

The most straightforward way to compute \rwc is via Monte Carlo sampling.
We use this approach in an earlier version of this work~\citep{garimella2016quantifying}, with samples of \num{10000} random walks.
Nevertheless, collecting a large number of samples is computationally intensive, and leads to slow evaluation of \rwc.
% Moreover, it is only applicable on undirected graphs.
In this section, we propose a variant of \rwc defined as a special case of a \emph{random walk with restart} -- thus leading to a much more efficient computation.
This variant can handle cases where the random walker gets stuck (i.e., dangling vertices), by using restarts.
This feature is important for two reasons: ($i$) retweet graphs (one of our main considerations in this paper) are inherently directed, hence the direction of endorsement should be taken into account, and ($ii$) since these directed graphs are very often star-like, there are a few authoritative users who generate information that spreads through the graph.
Our previous Monte Carlo sampling does not take into consideration such graph structure, and the direction of information propagation, as the random walk process needs to be made ergodic for the sampling process to function.
%We use personalized page rank to simulate a random walk with restart (RWR). To simulate the measure we have in the WSDM paper but to extend it for the case of directed, weighted graphs.

To define the proposed variant of \rwc, we assume there are two sides for a controversy, defined as two disjoint sets of vertices $X$ and $Y$.
In the original definition of the measure, we start multiple random walks from random vertices on either side, which  terminate once  they reach a high-degree vertex.
For this variant of \rwc, random walks do not terminate, rather they \emph{restart} once they reach a high-degree vertex.

More formally, we consider two instances of a random walk with restart ({RWR}), based on whether they start (and restart) from $X$ ($\mathrm{start} = X$) or $Y$ ($\mathrm{start} = Y$).
When $\mathrm{start} = X$, the RWR has a restart vector uniformly distributed over $X$, and zero for vertices in $Y$ (the situation is symmetric for $\mathrm{start} = Y$).
Moreover, the random walk runs on a modified graph with all outgoing edges from high-degree vertices removed.
This modification transforms the high-degree vertices into dangling vertices, hence forcing the random walk to restart once it reaches one of these vertices.\footnote{To compute the stationary distribution of the random walks, we use the implementation of Personalized PageRank from NetworkX \url{https://networkx.github.io/documentation/latest/reference/generated/networkx.algorithms.link_analysis.pagerank_alg.pagerank.html}.}

To formally define this variant of \rwc, let $P_1$ and $P_2$ be the stationary distributions of the RWR obtained for $\mathrm{start} = X$ and $\mathrm{start} = Y$, respectively. We consider the conditional probability $\prob[\mathrm{start}=A \mid \mathrm{end}=B^+]$ that the random walk had started on side $A\in\{X, Y\}$, given that at some step at steady-state it is found in one of the high-degree vertices of side $B\in\{X, Y\}$ (denoted as $B^+$).
We thus consider the following four probabilities:
\begin{equation}
P_{_{X,X^+}} = \prob[\mathrm{start}=X \mid \mathrm{end}=X^+] = \frac{\frac{|X|}{|V|} \sum_{v \in X^+}P_1(v)}{\frac{|X|}{|V|}   \sum_{v \in X^+}P_1(v) + \frac{|Y|}{|V|}   \sum_{v \in X^+}P_2(v)},
\label{eq:eq1}
\end{equation}
\begin{equation}
P_{_{X,Y^+}} = \prob[\mathrm{start}=X \mid \mathrm{end}=Y^+] = \frac{\frac{|X|}{|V|} \sum_{v \in Y^+}P_1(v)}{\frac{|X|}{|V|}   \sum_{v \in Y^+}P_1(v) + \frac{|Y|}{|V|}   \sum_{v \in Y^+}P_2(v)},
\label{eq:eq2}
\end{equation}
\begin{equation}
P_{_{Y,Y^+}} = \prob[\mathrm{start}=Y \mid \mathrm{end}=Y^+] = \frac{\frac{|Y|}{|V|} \sum_{v \in Y^+}P_2(v)}{\frac{|X|}{|V|}   \sum_{v \in Y^+}P_1(v) + \frac{|Y|}{|V|}   \sum_{v \in Y^+}P_2(v)},
\label{eq:eq3}
\end{equation}
\begin{equation}
P_{_{Y,X^+}} = \prob[\mathrm{start}=Y \mid \mathrm{end}=X^+] = \frac{\frac{|Y|}{|V|} \sum_{v \in X^+}P_2(v)}{\frac{|X|}{|V|}   \sum_{v \in X^+}P_1(v) + \frac{|Y|}{|V|} \sum_{v \in X^+}P_2(v)}.
\label{eq:eq4}
\end{equation}
Notice that for the probabilities above we have
\begin{equation*}
	\prob[\mathrm{start}=X \mid \mathrm{end}=X^+] + \prob[\mathrm{start}=Y \mid \mathrm{end}=X^+] = 1
\end{equation*}
and
\begin{equation*}
	\prob[\mathrm{start}=X \mid \mathrm{end}=Y^+] + \prob[\mathrm{start}=Y \mid \mathrm{end}=Y^+] = 1
\end{equation*}
as we ought to.
The variant of the \rwc score can be now defined as
\begin{equation}
	\rwc = P_{_{X X^+}}  P_{_{Y Y^+}} - P_{_{X Y^+}}  P_{_{Y X^+}},
\end{equation}
which, like the original version, intuitively captures the difference in the probability of staying on the same side and crossing the boundary.

%And it works! (and faster). See Table~\ref{tab:new_randomwalk_score}.

\begin{figure}[t]
\centering
\includegraphics[width=0.5\textwidth]{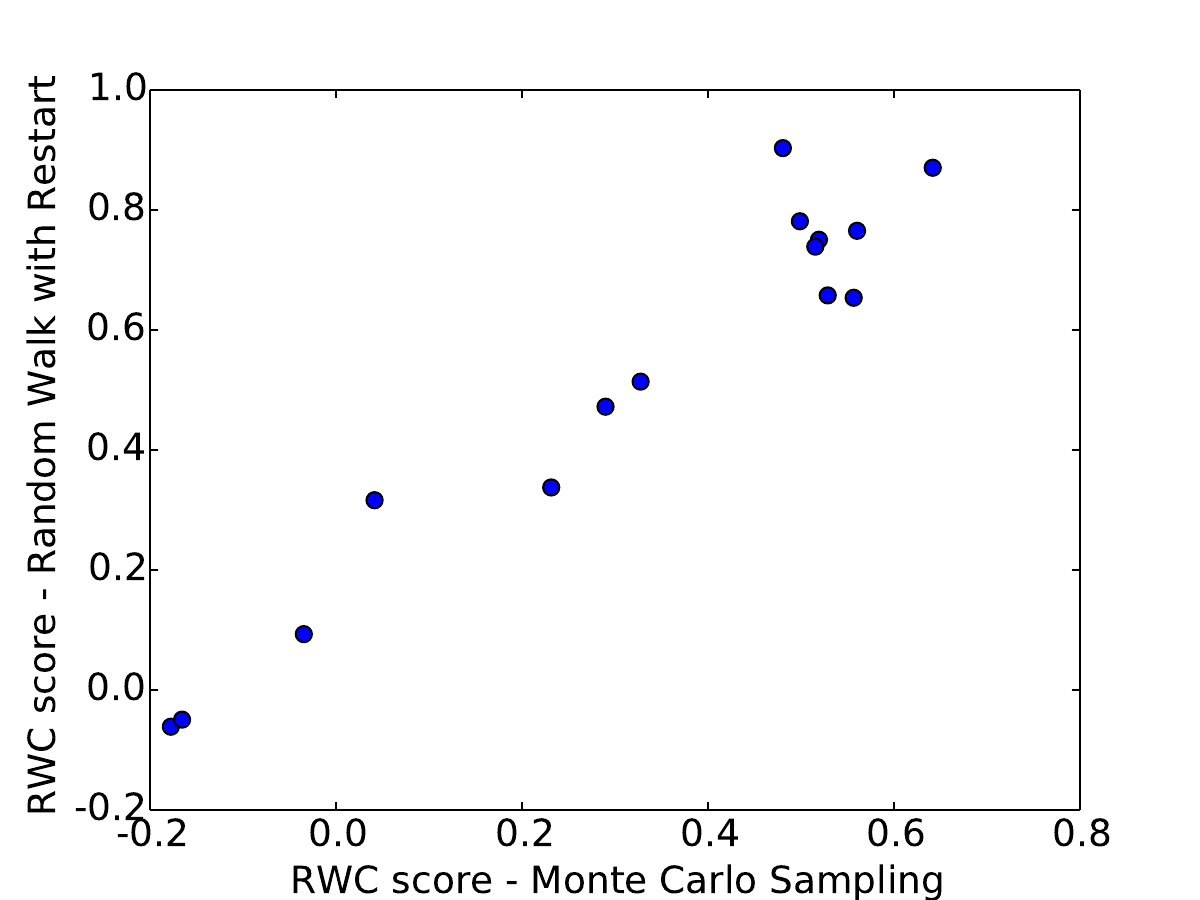}
\caption{Comparison between \rwc scores computed via Monte Carlo sampling and those computed via RWR. Pearson's $r$ = 0.96.}
\label{fig:new_randomwalk_score}
%\note{Axis labels are small, and should be more explicit in what they represent.}
%\vspace{-\baselineskip}
\end{figure}

To verify that the new variant of the score works as expected, we compare it to the original version of the score (obtained via Monte Carlo sampling).
The results are shown in Figure~\ref{fig:new_randomwalk_score},
from which it can be clearly seen that the new variant is almost identical to the original one. 
However, for the datasets considered in this work, we found empirically that this algorithm based on random walk with restart is up to $200$ times faster compared to the original Monte Carlo algorithm.

%\begin{table}
%\centering
%\begin{tabular}{c|c|c}
%hashtag & old score & new score \\
%\hline
%beeban & 0.6805062179 & 0.903138093335 \\
%nemtsov & 0.7287441311 & 0.657685209825 \\
%netanyahu & 0.5275613591 & 0.513954281506 \\
%russia\_march & 0.8416492317 & 0.870402361406 \\
%indiasdaughter & 0.7192917448 & 0.750407915721 \\
%baltimoreriots & 0.6987998005 & 0.781222823723 \\
%indiana & 0.7602860921 & 0.765360552532 \\
%ukraine & 0.7566430271 & 0.653736278535 \\
%leadersdebate & 0.7153758514 & 0.73892092953 \\
%sxsw & 0.4898069118 & 0.472180161452 \\
%1dfamheretostay & 0.4314132728 & 0.337591946046 \\
%germanwings & 0.1654044558 & 0.0930839395075 \\
%ultralive & 0.2414760332 & 0.316312108152 \\
%mothersday & 0.0224435961 &  -0.061158782692\\
%nepal & 0.0344725383 & -0.0493954802778 \\
%\end{tabular}
%\caption{Comparison between old and new randomwalk scores}
%\label{tab:new_randomwalk_score}
%\end{table}

\subsection{Betweenness}
\label{sec:betweenness}
Let us consider the set of edges $C \subseteq E$ in the cut defined by the
two partitions $X,Y$. This measure uses the notion of edge betweenness
and how the betweenness of the cut differs from that of the other edges.
Note that the cut here refers to the partioning obtained using METIS, as described in Section~\ref{sec:pipeline}.
Recall that the betweenness centrality $\bc( e)$ of an edge $e$ is defined as

\begin{equation}
	\bc( e) = \sum_{s \neq t \in V} \frac{\sigma_{s,t}(e)}{\sigma_{s,t}},
\end{equation}
where $\sigma_{s,t}$ is the total number of shortest paths between 
vertices $s,t$ in the graph and $\sigma_{s,t}(e)$ is the number of those 
shortest paths that include edge $e$.

The intuition here is that, if the two partitions are well-separated, then
the cut will consist of edges that bridge \emph{structural holes}~\cite{burt2009structural}.
In this case, the shortest paths that connect vertices of the two partitions will pass through the edges
in the cut, leading to high betweenness values for edges in $C$.
On the other hand, if the two partitions are not well separated, then the cut will
consist of \emph{strong ties}. In this case, the paths
that connect vertices across the two partitions will pass through one of the many edges
in the cut, leading to betweenness values for $C$ similar to the rest of the graph.

Given the distributions of edge betweenness on the cut and the rest of the graph, we compute the KL divergence $d_{_{KL}}$ of the two distributions by using kernel density estimation to compute the PDF and sampling \num{10000} points from each of these distributions (with replacement).
We define the \texttt{Betweenness Centrality Controversy} ($\mathit{BCC}$) measure as
\begin{equation}
	\mathit{BCC} = 1 - e^{-d_{_\mathit{KL}}},
\end{equation}
which assumes values close to zero when the divergence is small, and close to one when the divergence is large.

\subsection{Embedding}
\label{sec:embedding}
This measure is based on a low-dimensional embedding of graph $G$
produced by Gephi's ForceAtlas2 algorithm~\cite{jacomy2014forceatlas2}
(the same algorithm used to produce the plots in Figures~\ref{fig:visualization} and~\ref{fig:codicil}).
%We opt for this algorithm as it produces well-separated plots for controversial topics.
According to~\citet{noack2009modularity}, a force-directed embedding also maximizes modularity. Based on this observation, the two-dimensional layouts produced by this algorithm indicate a layout with maximum modularity.
%	We should add some of those ideas here to support our case.

Let us consider the two-dimensional embedding $\phi(v)$ of
vertices $v\in V$ produced by ForceAtlas2.
Given the partition $X$, $Y$ produced by the second stage of the pipeline,
we calculate the following quantities:
\begin{squishlist}
	\item $d_{_{X}}$ and $d_{_{Y}}$, the average embedded distance among pairs of vertices in the same partition, $X$ and $Y$ respectively;
	\item $d_{_{XY}}$, the average embedded distance among pairs of vertices
	across the two partitions $X$ and $Y$.
\end{squishlist}
Inpsired by the Davies-Bouldin (DB) index~\cite{daviesbouldin},
we define the {\tt Embedding Controversy} measure $\mathit{EC}$ as
\begin{equation}
	\mathit{EC} = 1 - \frac{d_{_{X}} + d_{_{Y}}}{2d_{_{XY}}}.
\end{equation}

$\mathit{EC}$ is close to one for controversial topics, corresponding to better-separated graphs and thus to higher degree of controversy, and close to zero for non-controversial topics.
%The measure takes values between zero and one, with values closer to one
%corresponding to better-separated graphs and thus to higher degree of controversy.

\subsection{Boundary Connectivity}
\label{sec:boundary}
This controversy measure was proposed by~\citet{guerra2013measure},
and is based on the notion of boundary and internal vertices.
Let $u\in X$ be
a vertex in partition $X$; $u$ belongs to the \emph{boundary} of
$X$ iff it is connected to at least one vertex of the other partition $Y$, and it is connected to at least one vertex
in partition $X$ that is not connected to any vertex of partition $Y$.
Following this definition, let $B_{_X},B_{_Y}$ be the set of boundary
vertices for each partition, and $B = B_{_X} \cup B_{_Y}$ the set of all boundary vertices.
By contrast, vertices $I_{_X} = X - B_{_X}$ are said to be the
\emph{internal} vertices of partition $X$ (similarly for $I_{_Y}$).
Let $I = I_{_X} \cup I_{_Y}$ be all internal vertices in either partition.
The reasoning for this measure is that, if the two partitions
represent two sides of a controversy, then boundary vertices will
be more strongly connected to internal vertices than to other boundary 
vertices of either partition.
This intuition is captured in the formula
\begin{equation}
	\mathit{GMCK} = \frac{1}{|B|}\sum_{u\in B}\frac{d_i(u)}{d_b(u) + d_i(u)} - 0.5 ,
\end{equation}
where $d_i(u)$ is the number of edges between vertex $u$ and internal
vertices $I$, while $d_b(u)$ is the number of edges between vertex $u$
and boundary vertices $B$.
% Note that this formula differs slightly from the original one
% in~\cite{guerra2013measure} in that it uses a different offset
% to take values between zero and one, like the other measures we evaluate.
Higher values of the measure correspond to higher degrees of controversy.

% We could get access to the datasets that were used in their paper, but slightly different in size. We replicated the results on what was available.
% Political blogs data from~\cite{adamic2005political} and Karate club.
% Results in Table~\ref{tab:external_datasets}

\subsection{Dipole Moment}
\label{sec:dipole}
This controversy measure was presented by~\citet{morales2015measuring},
and is based on the notion of \emph{dipole moment} that has its origin in physics.
Let $R(u)\in [-1,1]$ be a polarization value assigned to vertex $u\in V$.
Intuitively, extreme values of $R$ (close to $-1$ or $1$) correspond
to users who belong most clearly to either side of the controversy.
To set the values $R(u)$ we follow the process described
in the original paper~\cite{morales2015measuring}: we set $R = \pm 1$ for the top-$5\%$
highest-degree vertices in each partition $X$ and $Y$, and set the
values for the rest of the vertices by label-propagation.
Let $n^+$ and $n^-$ be the number of vertices $V$ with
positive and negative polarization values, respectively, and $\Delta A$ the
absolute difference of their normalized size
$ %\begin{equation*}
	\Delta A = \left|\frac{ n^+  - n^- }{|V|} \right| .
$ %\end{equation*}
Moreover, let $gc^+$ ($gc^-$) be the average polarization value among
vertices $n^+$ ($n^-$) and set $d$ as half their absolute difference,
$ %\begin{equation*}
	d = \frac{ \left| gc^+ - gc^- \right| }{2}.
$ %\end{equation*}
The dipole moment controversy measure is defined as
\begin{equation}
	\mathit{MBLB} = (1 - \Delta A)d.
\end{equation}
The rationale for this measure is that, if the two partitions
$X$ and $Y$ are well separated, then label propagation will assign different extreme
($\pm1$) $R$-values to the two partitions, leading
to higher values of the $\mathit{MBLB}$ measure. Note also that larger
differences in the size of the two partitions (reflected in the
value of $\Delta A$) lead to decreased values for the measure,
which takes values between zero and one.

%One drawback of this measure is that it depends significantly
%on label propagation for the assignment of $R$-values and
%thus on the seed labeling of the vertices.

% Drawback: Need to pick labelled seeds to start the label propagation. controversy score depends on seeds and the number of seeds picked and the results differ significantly (see Figures~\ref{pol_retweet2} (a and b)).

% Figure~\ref{fig:venezuela_timeline}.

% \todo[Kiran]{add more}

%%%%%%%%%%%%%%%%%%%%%%%%%%%%%%%%%%%%%%%%%%%%%%%%%%%%%%%%%%%%%%%%%%

%%%%%%%%%%%%%%%%%%%%%%%%%%%%%%%%%%%%%%%%%%%%%%%%%%%%%%%%%%%%%%%%%%

\section{Controversy scores for users}
\label{sec:user-controversy}

The previous sections present measures to quantify the controversy of a conversation graph.
In this section, we propose two measures to quantify the controversy of a single user in the graph.
We denote this score as a real number that takes values in $[-1, 1]$, with $0$ representing a neutral score, and $\pm1$ representing the extremes for each side.
Intuitively, the controversy score of a user indicates how `biased' the user is towards a particular side on a topic. For instance, for the topic `abortion', pro-choice/pro-life activist groups tweeting consistently about abortion would get a score close to -1/+1 while average users who interact with both sides would get a score close to zero. In terms of the positions of users on the retweet graph, a neutral user would lie in the `middle', retweeting both sides, where as a user with a high controversy score lies exclusively on one side of the graph.

\spara{\textbf{$RWC^{user}$:}}
The first proposed measure is an adaptation of \rwc. 
As input, we are given a user $u\in V$ in the graph and a partitioning of the graph into two sides, defined as disjoint sets of vertices $X$ and $Y$. 
We then consider a random walk that starts -- and restarts -- at the given user u.
Moreover, as with \rwc, the high-degree vertices on each side ($X^+$ and $Y^+$) are treated as dangling vertices -- whenever the random walk reaches these vertices, it teleports to vertex $u$ with probability $1$ in the next step.
To quantify the controversy of $u$, we ask how often the random walk is found on vertices that belong to either side of the controversy.
Specifically, for each user $u$, we consider the conditional probabilities 
$\prob[\mathrm{start} =u \mid  \mathrm{end} =X^+]$ and $\prob[\mathrm{start} =u \mid \mathrm{end} =Y^+]$ , and estimate them by using the power iteration method. 
% We initialize the power iteration with a vector which contains a 1 for the user $u$ and zeros for all other users. The restart vector is also set to this. After the power iteration finishes, we just only consider the probabilities from the stationery distribution corresponding to the high-degree vertices to compute the above probabilities.
Assuming that user $u$ belongs to side $X$ of the controversy (i.e., $u \in X$), their controversy is defined as:
\begin{equation}
RWC^{user}(u,X) = \frac{\prob[\mathrm{start} =u \mid \mathrm{end} =X^+]}{\prob[\mathrm{start} =u \mid  \mathrm{end} =X^+] + \prob[\mathrm{start} =u \mid  \mathrm{end} =Y^+]} .
\end{equation}

\spara{Expected hitting time:}
The second proposed measure is also random-walk-based, but defined on the expected number of steps to hit the high-degree vertices on either side.
Intuitively, a vertex is assigned a score of higher absolute value (closer to $1$ or $-1$), if, compared to other vertices in the graph, it takes a very different time to reach a high-degree vertex on either side ($X^+$ or $Y^+$).
Specifically, for each vertex $u\in V$ in the graph, we consider a random walk that starts at $u$, and estimate the expected number of steps, $l_u^{_X}$  before the random walk reaches any high-degree vertex in $X^+$.
Considering the distribution of values of $l_u^{_X}$ across all vertices $u \in V$, we define $\rho^{_X}(u)$ as the fraction of vertices $v\in V$ with $l_v^{_X} < l_u^{_X}$.
We define $\rho^{_Y}(u)$ similarly.
Obviously, we have $\rho^{_X}(u), \rho^{_Y}(u) \in [0,1)$.
The controversy score of a user is then defined as
\begin{equation}
	\rho(u) = \rho^{_X}(u) - \rho^{_Y}(u) \in (-1, 1) .
\end{equation}
Following this definition, a vertex that, compared to most other vertices, is very close to high-degree vertices $X^+$ will have $\rho^{_X}(u) \approx 1$; and if the same vertex is very far from high-degree vertices $Y^+$, it will have $\rho^{_Y}(u) \approx 0$ -- leading to a controversy score $\rho(u) \approx 1 - 0 = 1$.
The opposite is true for vertices that are far from $X^+$ but close to $Y^+$, leading to a controversy score $\rho(u) \approx -1$.

% We also have another version of the controversy score which works better. It is based on the expected number of steps for the randomwalk starting from a user to hit a high-degree vertex on either side.

% It works as follows: (i) starting from each vertex, perform random walks for a large number of times. Restart the randomwalk when you hit a high-degree vertex on either side. 
% (ii) Count the number of steps taken to reach the high-degree vertex separately for each side.
% (iii) For each side, consider the mean of the number of steps to reach a high-degree vertex for each of the random walks as the expected number of steps taken for each user.
% (iv) For each vertex, for each side, compute a distribution of the number of steps taken and divide the distribution into 100 buckets, based on percentiles.
% (v) For each vertex, for each side, compute the bucket to which the vertex belongs to on that side.
% (vi) The controversy score of that vertex is defined as ($bucket\_id\_side_1 - bucket\_id\_side_2)/100$, which gives a score between -1 and 1.

\subsection{Comparison with BiasWatch}

BiasWatch~\cite{lu2015biaswatch} is a recently-proposed, light-weight approach to compute controversy scores for users on Twitter. 
At a high level, the BiasWatch approach consists of the following steps:
\begin{enumerate}
	\item Hand pick a small set of seed hashtags to characterize the two sides of a controversy (e.g., \#prochoice vs. \#prolife);
	\item Expand the seed set of hashtags based on co-occurrence;
	\item Use the two sets of hashtags, identify strong partisans in the graph (users with high controversy score);
	\item Assign controversy scores to other users via a simple label propagation approach. %The proposed label propagation approach is defined in a way to minimize propagation of noisy labeling.
\end{enumerate}

\begin{figure}[!ht]

    \begin{tabular}[b]{l c}
    \toprule
      Topic & Pearson correlation \\ 
      \midrule
	\#abortion &  0.51 \\
	\#obamacare &  0.48 \\
	\#guncontrol &  0.42 \\
	\#beefban &  0.41 \\
	\#baltimoreriots &  0.41 \\
	\#netanyahuspeech &  0.41 \\
	\#nemtsov &  0.38 \\
	\#indiana &  0.40 \\
	\#indiasdaughter &  0.39 \\
	\#ukraine & 0.40 \\
      \bottomrule
    \end{tabular}
    \centering
	\includegraphics[width=0.48\textwidth]{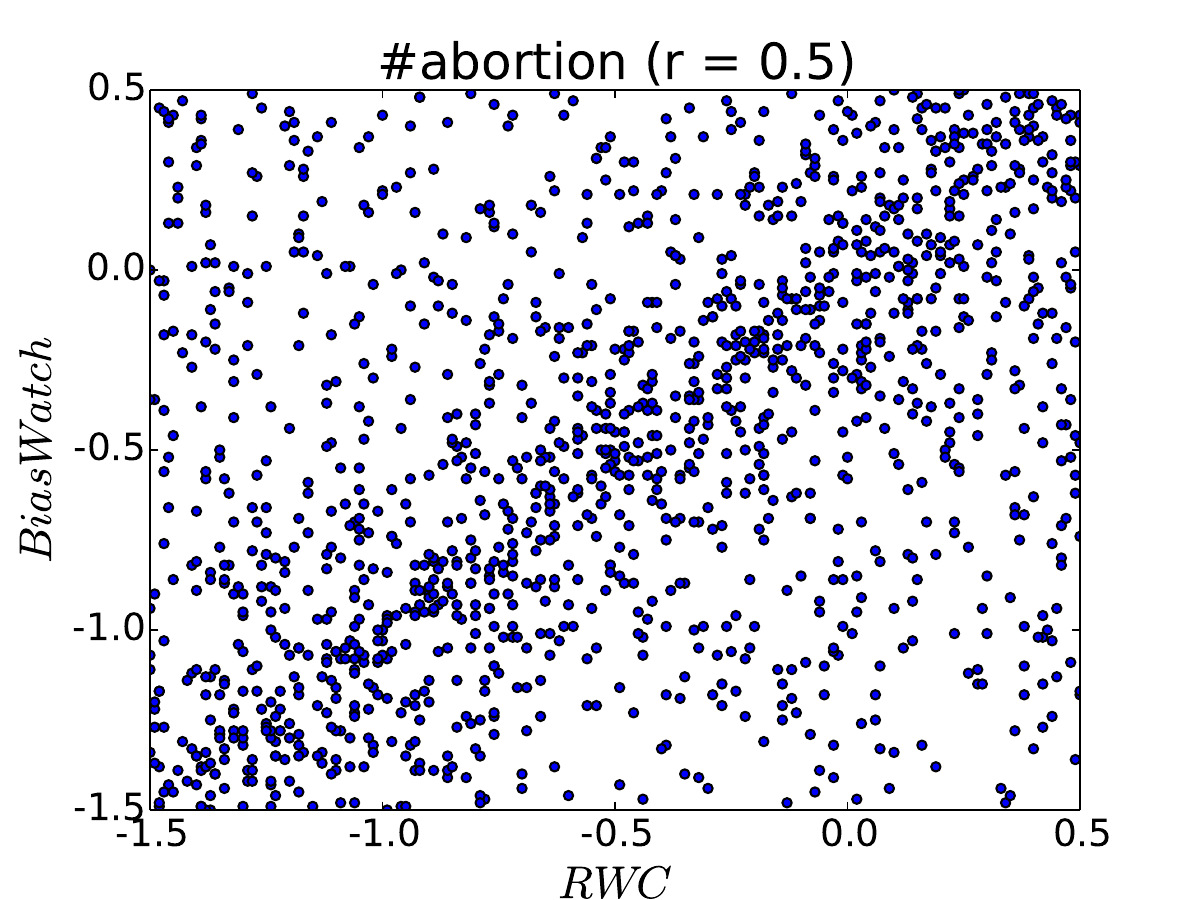}
    \hfill
%    \captionlistentry[table]{A table beside a figure}
%    \captionsetup{labelformat=andtable}
	\caption{(left) Pearson's $r$ between the scores obtained by our algorithm (Expected hitting time) and BiasWatch. (right) Sample scatter plot for \#abortion.}
	\label{fig:pearsonrbiaswatch}
\end{figure}

We compare the controversy scores obtained by our approaches to the ones obtained by BiasWatach\footnote{For BiasWatch we use parameters $\mu_1 = 0.1$, $\mu_2 = 0.4$, optimization method `COBYLA', cosine similarity threshold $0.4$, and $10$ nearest neighbors for hashtag extension.}
on two sets of datasets: tweets matching the hashtags ($i$) \#obamacare, \#guncontrol, and \#abortion, provided by \citet{lu2015biaswatch} and ($ii$) the datasets in Table~\ref{tab:datasets}.
We compute the Pearson correlation between our measure based on Expected hitting time and BiasWatch; the results are shown in Figure~\ref{fig:pearsonrbiaswatch}.
We omit the comparison with $RWC^{user}$ scores as they are almost identical to the ones by BiasWatch.

%\note[Michael]{What approach do we use to compare with BiasWatch? (I assume it's the one based on expected length?)}

The authors also provide datasets which contain human annotations for controversy score (in the range [-2,2]) for 500 randomly selected users. We discretize our controversy scores to the same range, and compute the 5-category Fleiss' $\kappa$ value. The $\kappa$ value is 0.35, which represents a `fair' level of agreement, according to~\citet{landis1977measurement}.

%Then, as they do in the Biaswatch paper, I discretized the outputs of both our and their algorithm to a range of [-2,2] and computed the 5-category Fleiss $\kappa$ value, which was around 0.35, which is decent according to their paper.

Our approach thus provides results that are similar to the state-of-the-art approach.
%, which encourages us to further explore how to apply this random-walk-based framework we propose to other tasks.
Our method also has two advantages over the BiasWatch measure: (i) Even though we do not make use of any content information in our measure, we perform at par; and (ii) $RWC^{user}$ provides an intuitive extension to our RWC measure. Given this unified framework, it is possible to design ways to reduce controversy, e.g., by connecting opposing views~\cite{garimella2017reducing,garimella2017factors}, and such a unified formulation can help us define principled objective functions to approach these tasks.
%, such our unified frame We do not make use of content
% Our objective here is not to design yet another algorithm to provide controversy scores for users, but to show the applicability of our random walk based score as a single framework for a wide range of tasks.

%} % end color = red

%%%%%%%%%%%%%%%%%%%%%%%%%%%%%%%%%%%%%%%%%%%%%%%%%%%%%%%%%%%%%%%%%%

%%%%%%%%%%%%%%%%%%%%%%%%%%%%%%%%%%%%%%%%%%%%%%%%%%%%%%%%%%%%%%%%%%

\section{Experiments}
\label{sec:experiments}
In this section we report the results of the various configurations of the pipeline proposed in this paper.
As previously stated, we omit results for the content and hybrid graph building approaches presented in Section~\ref{sec:graph_building}, as they do not perform well.
We instead focus on the retweet and follow graphs, and test all the measures presented in Section~\ref{sec:measure} on the topics described in Table~\ref{tab:datasets}.
In addition, we test all the measures on a set of external datasets used in previous studies~\cite{adamic2005political, conover2011political, guerra2013measure} to validate the measures against a known ground truth.
Finally, we use an evolving dataset from Twitter collected around the death of Venezuelan president Hugo Chavez~\cite{morales2015measuring} to show the \emph{evolution} of the controversy measures in response to high-impact events.

To avoid potential overfitting, we use only eight graphs as testbed during the development of the measures, half of them controversial (beefban, nemtsov, netanyahu, russia\_march) and half non-controversial (sxsw, germanwings, onedirection, ultralive).
%We report the results of the controversy measures when applied to the remaining twelve graphs.
This procedure resembles a 40/60\% train/test split in traditional machine learning applications.\footnote{A demo of our controversy measures %on a large number of Twitter hashtags 
can be found at \url{https://users.ics.aalto.fi/kiran/controversy}.}

\begin{figure}
\centering
\begin{minipage}{0.45\textwidth}
\centering
%\exedout % first figure itself
\includegraphics[width=\textwidth, clip=true, trim=0 40 10 50]{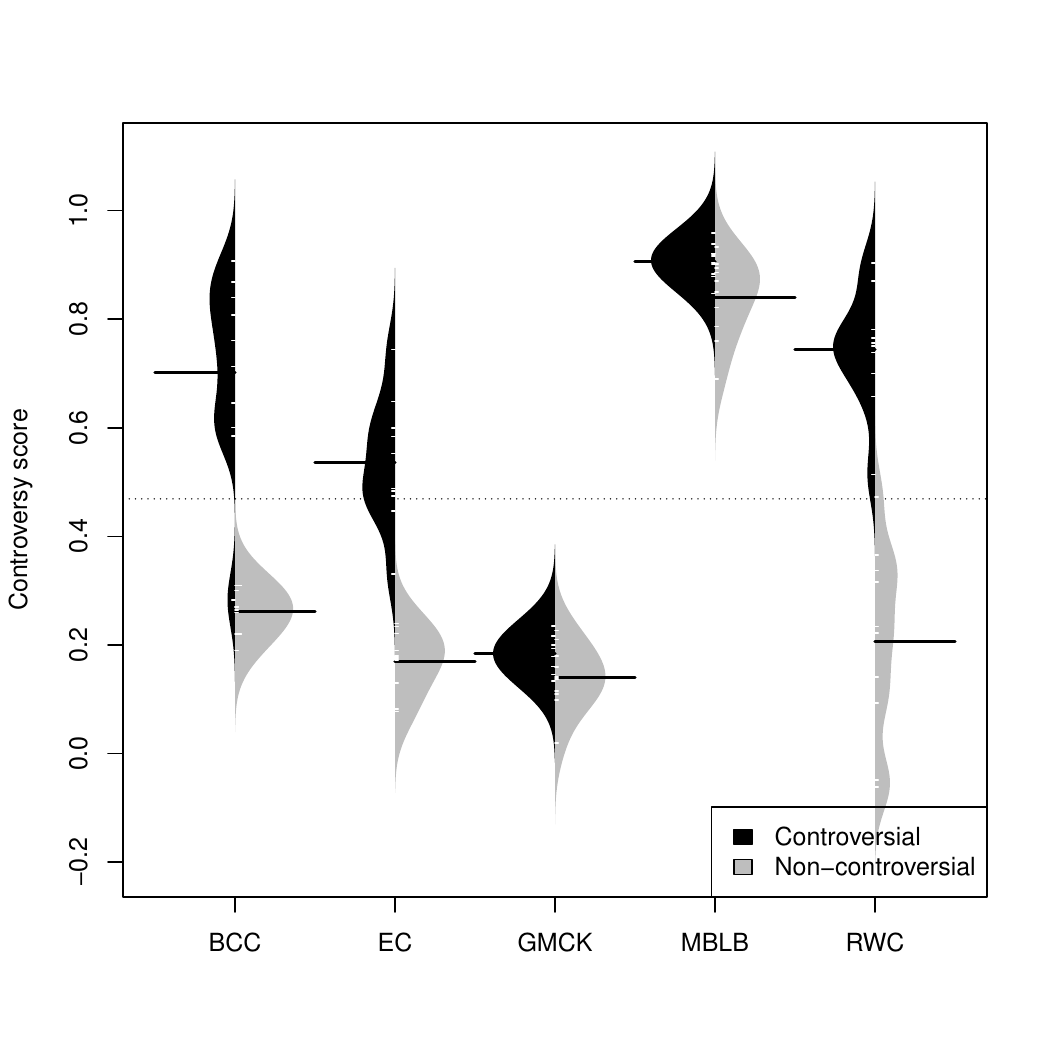}
\caption{Controversy scores on \emph{retweet} graphs of various controversial and non-controversial datasets.}
\label{fig:pol_retweet}
\end{minipage}\hfill
\begin{minipage}{0.45\textwidth}
\centering
%\exedout % second figure itself
\includegraphics[width=\textwidth, clip=true, trim=0 40 10 50]{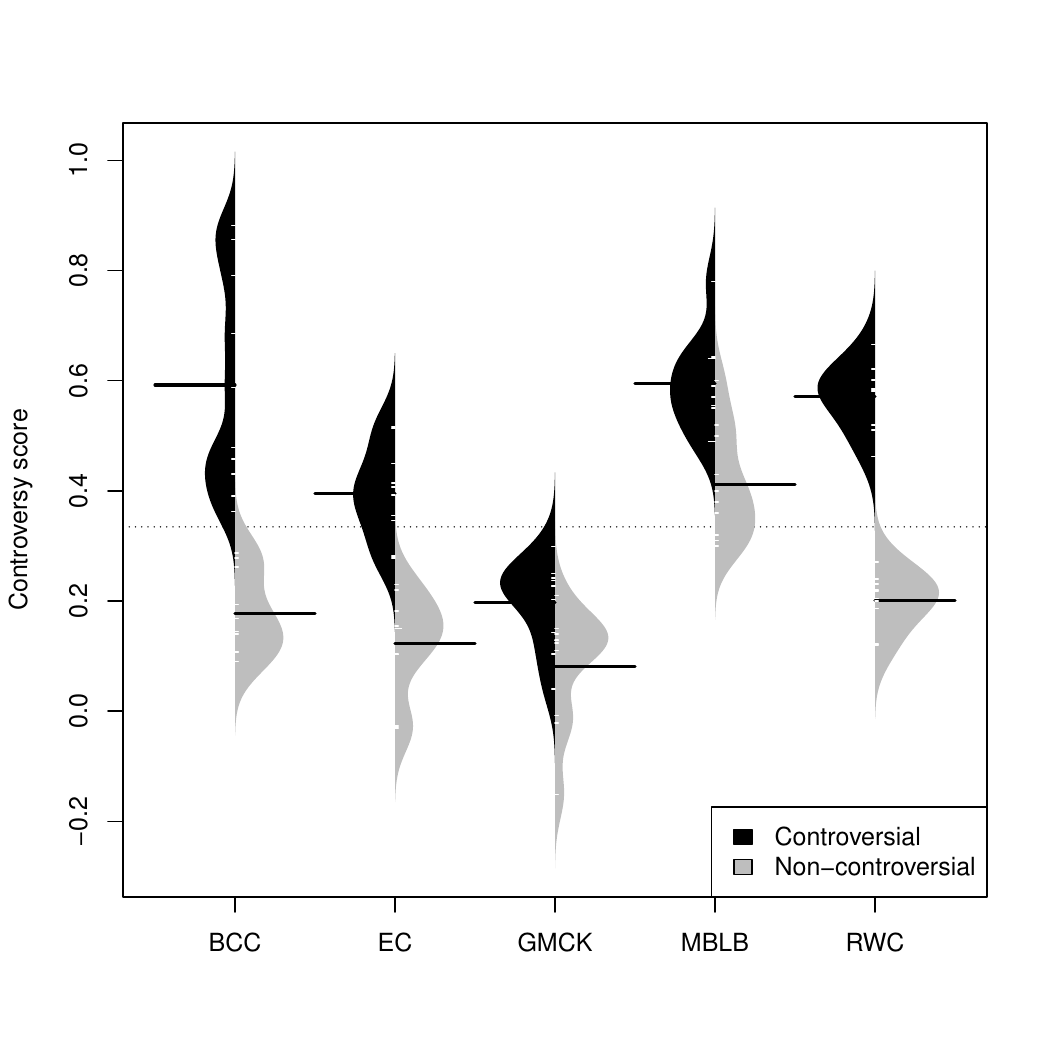}
\caption{Controversy scores on \emph{follow} graphs of various controversial and non-controversial datasets.}
\label{fig:pol_follow}
\end{minipage}
\end{figure}

%\begin{figure}[tb]
%\includegraphics[width=0.5\textwidth, clip=true, trim=0 40 10 50]{beanplot_retweet_graph}
%%\vspace{-\baselineskip}
%\end{figure}
%\begin{figure}[tb]
%\includegraphics[width=0.5\textwidth, clip=true, trim=0 40 10 50]{beanplot_follow_graph}
%%\vspace{-\baselineskip}
%\end{figure}

\begin{table*}[t]
\caption{Results on external datasets. The `C?' column indicates whether the previous study considered the dataset controversial (ground truth).}
\label{tab:external_datasets}
\centering
\small
\begin{tabular}{l r r r r r r r r}
\toprule
Dataset & $|V|$ & $|E|$ & C? & $\rwc$ & $\mathit{BCC}$ & $\mathit{EC}$ & $\mathit{GMCK}$ & $\mathit{MBLB}$ \\
\midrule
Political blogs		& \num{1222} & \num{16714} & \cmark	& \num{0.42} & \num{0.53} & \num{0.49} & \num{0.18} & \num{0.45} \\
Twitter politics	& \num{18470} & \num{48053} & \cmark	& \num{0.77} & \num{0.79} & \num{0.62} & \num{0.28} & \num{0.34} \\
Gun control		& \num{33254} &	\num{349782} & \cmark	& \num{0.70} & \num{0.68} & \num{0.55} & \num{0.24} & \num{0.81}\\
Brazil soccer		& \num{20594} & \num{82421} & \cmark	& \num{0.67} & \num{0.48} & \num{0.68} & \num{0.17} & \num{0.75} \\
Karate club		& \num{34} & \num{78} & \cmark	& \num{0.11} & \num{0.64} & \num{0.51} & \num{0.17} & \num{0.11} \\
Facebook university	& \num{281} & \num{4389} & \xmark	& \num{0.35} & \num{0.26} & \num{0.38} & \num{0.01} & \num{0.27}\\
NYC teams		& \num{95924} & \num{176249} & \xmark	& \num{0.34} & \num{0.24} & \num{0.17} & \num{0.01} & \num{0.19} \\
\bottomrule
\end{tabular}
%\vspace{-\baselineskip}
\end{table*}

\subsection{Twitter hashtags}
Figure~\ref{fig:pol_retweet} and Figure~\ref{fig:pol_follow} report the scores computed by each measure for each of the $20$ hashtags, on the retweet and follow graph, respectively.
Each figure shows a set of beanplots,\footnote{A beanplot is an alternative to the boxplot for visual comparison of univariate data among groups.} one for each measure.
Each beanplot shows the estimated probability density function for a measure computed on the topics, the individual observations are shown as small white lines in a one-dimensional scatter plot, and the median as a longer black line.
The beanplot is divided into two groups, one for controversial topics (left/dark) and one for non-controversial ones (right/light).
A larger separation of the two distributions indicates that the measure is better at capturing the characteristics of controversial topics.
For instance, this separation is fundamental when using the controversy score as a feature in a classification algorithm.

Figures~\ref{fig:pol_retweet} and~\ref{fig:pol_follow} clearly show that \rwc is the best measure on our datasets.
% Not only the two groups are well separated, but the cut point is around the mid-range point $0.5$ for the retweet graph (and around $0.4$ for the follow graph).
$\mathit{BCC}$ and $\mathit{EC}$ show varying degrees of separation and overlap, although $\mathit{EC}$ performs slightly better as the distributions are more concentrated, while $\mathit{BCC}$ has a very wide distribution.
The two baselines $\mathit{GMCK}$ and $\mathit{MBLB}$ instead fail to separate the two groups.
Especially on the retweet graph, the two groups are almost indistinguishable.

For all measures the median score of controversial topics is higher than for non-controversial ones.
This result suggests that both graph building methods, retweet and follow, are able to capture the difference between controversial and non-controversial topics.
Given the broad range of provenience of the topics covered by the dataset, and their different characteristics, the consistency of the results is very encouraging.

\subsection{External datasets}
We have shown that our approach works well on a number of datasets extracted in-the-wild from Twitter.
But, how well does it generalize to datasets from different domains?
%We experimentally answer this question in this section.

We obtain a comprehensive group of datasets kindly shared by authors of previous works:
\emph{Political blogs},
links between blogs discussing politics in the US~\cite{adamic2005political};
\emph{Twitter politics}, 
Twitter messages pertaining to the 2010 midterm election in US~\cite{conover2011political};
and the following five graphs used in the study that introduced $\mathit{GMCK}$~\cite{guerra2013measure},
(a) \emph{Gun control}, retweets about gun control after the shooting at the Sandy Hook school; %in Connecticut;
(b) \emph{Brazil soccer}, retweets about to two popular soccer teams in Brazil;
(c) \emph{Karate club}, the well-known social network by~\cite{zachary1977karate};
(d) \emph{Facebook university}, a social graph among students and professors at a Brazilian university;
(e) \emph{NYC teams}, retweets about two New York City sports teams.

Table~\ref{tab:external_datasets} shows a comparison of the controversy measures under study on the aforementioned datasets.\footnote{The datasets provided by \citet{guerra2013measure} are slightly different from the ones used in the original paper because of some irreproducible filtering used by the authors. We use the datasets provided to us verbatim.}
For each dataset we also report whether it was considered controversial in the original paper, which provides a sort of ``ground truth'' to evaluate the measures against.

All the measures are able to distinguish controversial graphs to some extent, in the sense that they return higher values for the controversial cases. The only exception is Karate club.
Both \rwc and $\mathit{MBLB}$ report low controversy scores for this graph.
It is possible that the graph is too small for such random-walk-based measures to function properly.
Conversely, $\mathit{BCC}$ is able to capture the desired behavior, which suggests that shortest-path and random-walk based measures might have a complementary function.

Interestingly, while the Political blogs datasets is often considered a gold standard for polarization and division in online political discussions, all the measures agree that it presents only a moderate level of controversy.
Conversely, the Twitter politics dataset is clearly one of the most controversial one across all measures.
This difference suggests that the measures are more geared towards capturing the dynamics of controversy as it unfolds on social media, which might differ from more traditional blogs.
For instance, one such difference is the \emph{cost} of an endorsement: placing a link on a blog post arguably consumes more mental resources than clicking on the retweet button.

For the `Gun control' dataset, \citeauthor{guerra2013measure} need to manually distinguish three different partitions in the graph: gun rights advocates, gun control supporters, and moderates.
Our pipeline is able to find the two communities with opposing views (grouping together gun control supporters and moderates, as suggested in the original study) without any external help.
All measures agree with the conclusions drawn in the original paper that this topic is highly controversial.

Note that even though from the results in Table~\ref{tab:external_datasets}, \rwc, $\mathit{BCC}$ and $\mathit{EC}$ appear to outperform each other, it is not the case. 
These methods are not comparable, meaning, a score of 0.5 for \rwc is not the same as a 0.5 for $\mathit{BCC}$.
The insight we can draw from these results is that our methods are able to discern a controversial topic from a non-controversial one consistently, irrespective of the domain, and are able to do so more reliably than existing methods ($\mathit{GMCK}$ and $\mathit{MBLB}$).

\begin{figure*}[t]
\includegraphics[width=\textwidth, clip=true, trim=0 25 0 30]{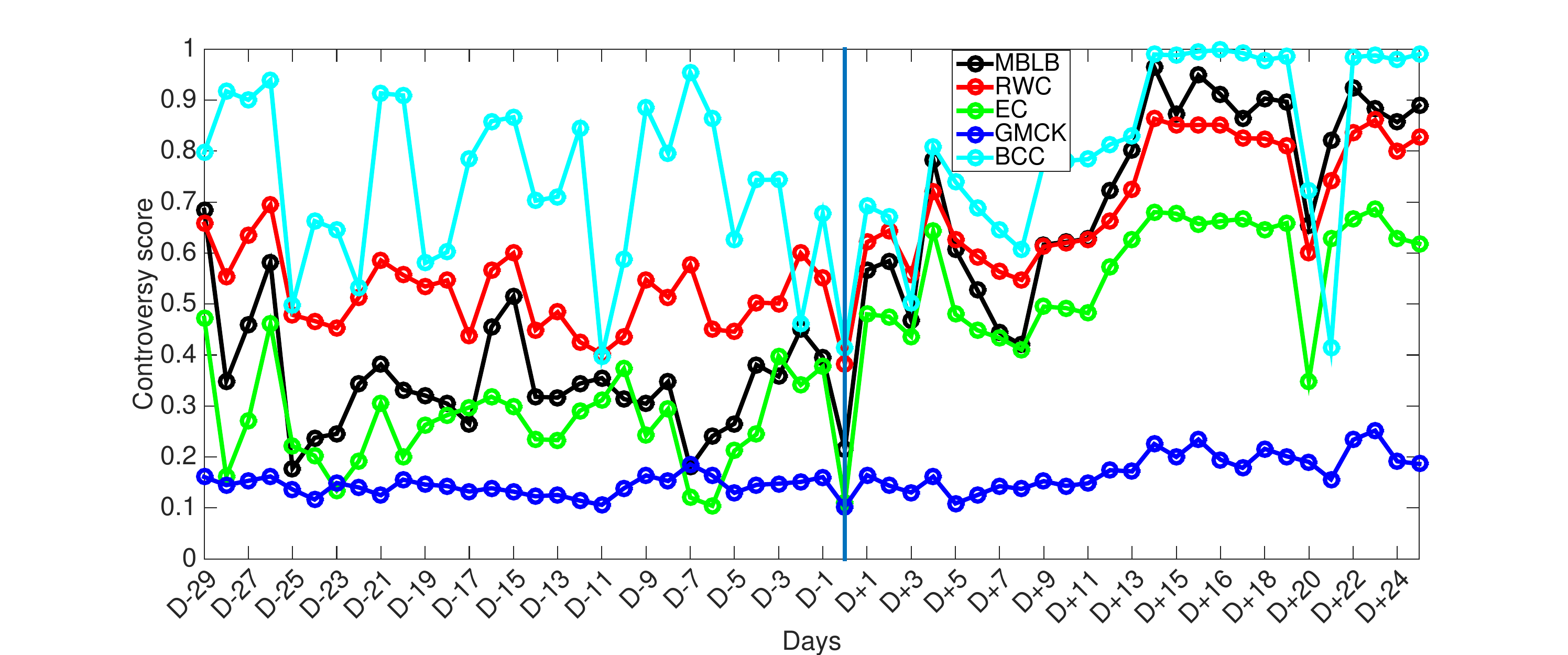}
\caption{Controversy scores on 56 retweet graphs from \citeauthor{morales2015measuring} Day `D' (indicated by the blue vertical line) indicates the announcement of the death of president Hugo Chavez. }
\label{fig:venezuela_timeline}
\vspace{-\baselineskip}
\end{figure*}

\subsection{Evolving controversy}
We have shown that our approach also generalizes well to datasets from different domains.
But in a real deployment the measures need to be computed continuously, as new data arrives.
How well does our method work in such a setting?
And how do the controversy measures evolve in response to high-impact events?

To answer these questions, we use a dataset from the study that introduced $\mathit{MBLB}$~\cite{morales2015measuring}.
The dataset comprises Twitter messages pertaining to political events in Venezuela around the time of the death of Hugo Chavez (Feb-May 2013).
The authors built a retweet graph for each of the 56 days around the day of the death (one graph per day).

Figure~\ref{fig:venezuela_timeline} shows how the intensity of controversy evolves according to the measures under study (which occurs on day `D').
The measure proposed in the original paper, $\mathit{MBLB}$, which we use as `ground truth', shows a clear decrease of controversy on the day of the death, followed by a progressive increase in the controversy of the conversation.
The original interpretation states that on the day of the death a large amount of people, also from other countries, retweeted news of the event, creating a single global community that got together at the shock of the news.
After the death, the ruling and opposition party entered in a fiery discussion over the next elections, which increased the controversy.

All the measures proposed in this work show the same trend as $\mathit{MBLB}$.
Both \rwc and $\mathit{EC}$ follow very closely the original measure (Pearson correlation coefficients $r$ of \num{0.944} and \num{0.949}, respectively), while $\mathit{BCC}$ shows a more jagged behavior in the first half of the plot ($r=\num{0.743}$), due to the discrete nature of shortest paths.
All measures however present a dip on day `D', an increase in controversy in the second half, and another dip on day `D+20'.
Conversely, $\mathit{GMCK}$ reports an almost constant moderate value of controversy during the whole period ($r=\num{0.542}$), with barely noticeable peaks and dips.
We conclude that our measures generalize well also to the case of evolving graphs, and behave as expected in response to high-impact events.

%{\color{blue}
\subsection{Simulations}
Given that \rwc is the best-performing score among the ones in this study, we focus our attention solely on it henceforth.
To measure the robustness of the \rwc score, we generate random Erd\"{o}s-R\'enyi graphs with varying community structure, and compute the \rwc score on them. 
Specifically, to mimic community structure, we plant two separate communities with intra-community edge probability $p_1$.
That is, $p_1$ defines how dense these communities are within themselves.
We then add random edges between these two communities with probability $p_2$.
Therefore, $p_2$ defines how connected the two communities are.
A higher value of $p_1$ and a lower value of $p_2$ create a clearer two-community structure.

\begin{figure}[t]
\begin{minipage}{.49\linewidth}
\centering
\subfloat[]{\label{fig:simulation1}\includegraphics[width=\textwidth]{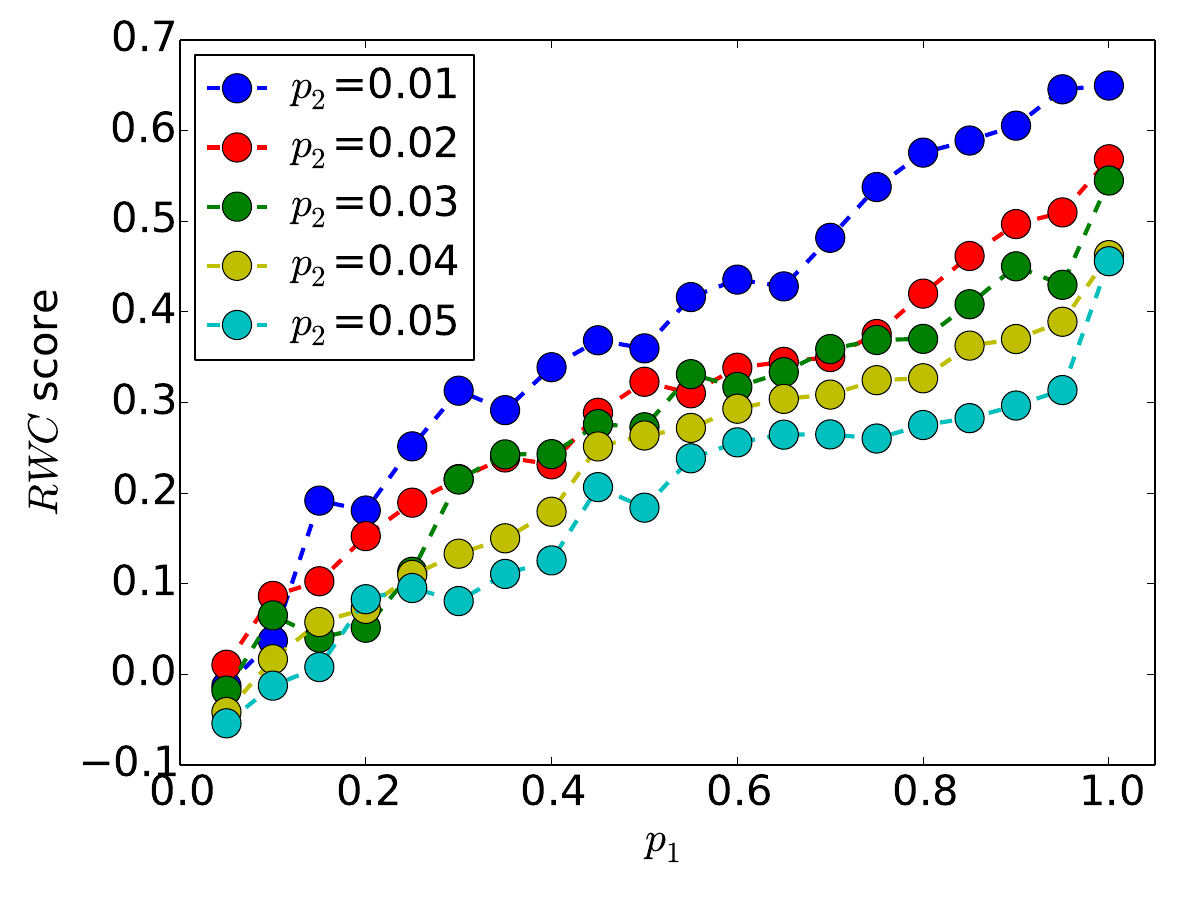}}
\end{minipage}%
\hspace{0.01\textwidth}
\begin{minipage}{.49\linewidth}
\centering
\subfloat[]{\label{fig:simulation2}\includegraphics[width=\textwidth]{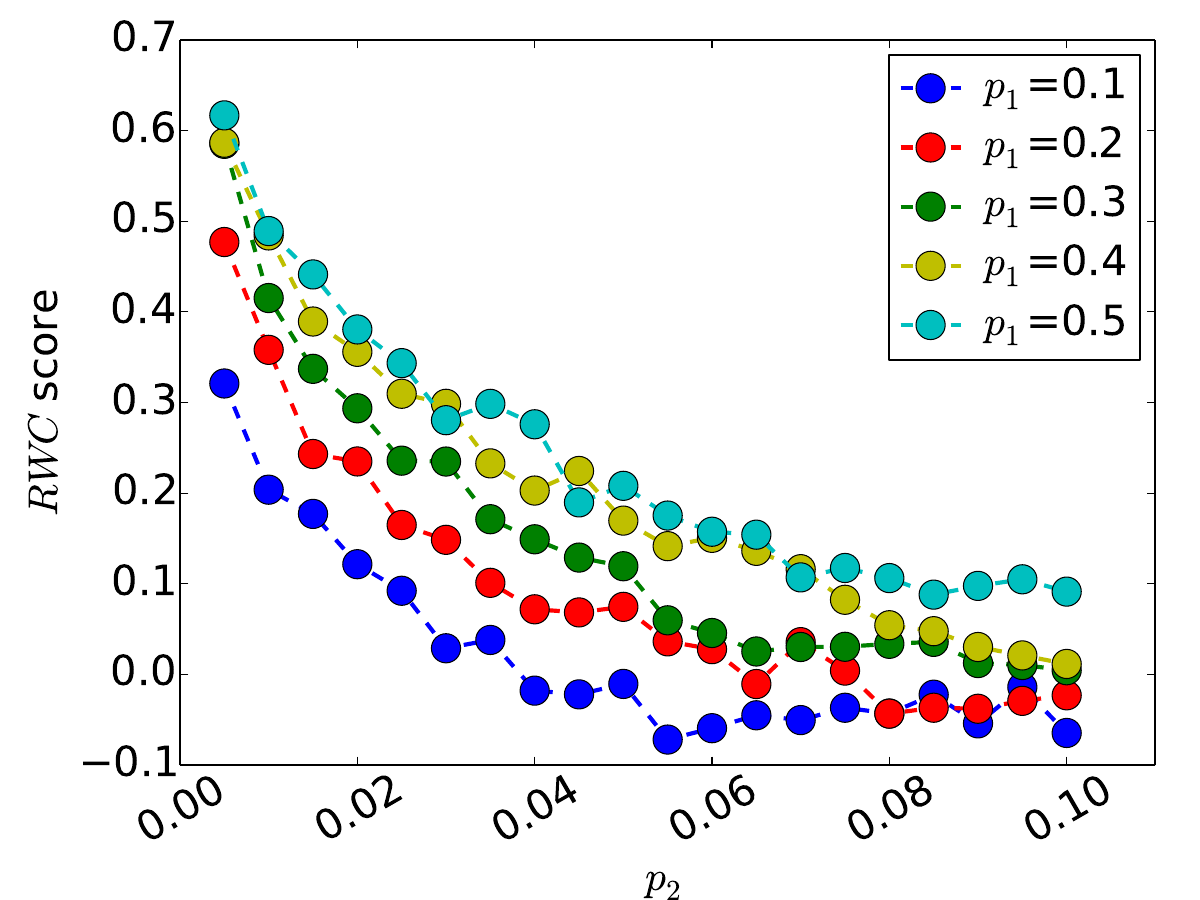}}
\end{minipage}\par\medskip
\caption{\rwc scores for synthetic Erd\"{o}s-R\'enyi graphs planted with two communities. $p_1$ is the intra-community edge probability, while $p_2$ is the inter-community edge probability.}
\label{fig:simulation}
\end{figure}

Figure~\ref{fig:simulation} shows the \rwc score for random graphs of \num{2000} vertices for two different settings: plotting the score as a function of $p_1$ while fixing $p_2$ (Figure~\ref{fig:simulation1}), and vice-versa (Figure~\ref{fig:simulation2}).
The \rwc score reported is the average over ten runs.
We observe a clear pattern:
the \rwc score increases as we increase the density within the communities,
and decreases as we add noise to the community structure.
The effects of the parameters is also expected, for a given value of $p_1$, a smaller value of $p_2$ generates a larger \rwc score, as the communities are more well separated.
Conversely, for a given value of $p_2$, a larger value of $p_1$ generates a larger \rwc scores, as the communities are denser.

%\begin{figure}
%\centering
%\begin{minipage}{0.48\textwidth}
%\centering
%%\exedout % first figure itself
%\includegraphics[width=\textwidth]{simulated_RWC1}
%\caption{fixed $p_2$, varying $p_1$}
%\label{fig:simulation1}
%\end{minipage}\hfill
%\begin{minipage}{0.48\textwidth}
%\centering
%%\exedout % second figure itself
%\includegraphics[width=\textwidth]{simulated_RWC2}
%\caption{fixed $p_1$, varying $p_2$}
%\label{fig:simulation2}
%\end{minipage}
%\end{figure}

\subsection{Controversy detection in the wild}
In most of the experiments presented so far, we hand-picked known topics which are controversial and showed that our method is able to separate them from the non-controversial topics.
To check whether our system works in a real-world setting, we deploy it in the wild to explore actual topics of discussion on Twitter and detect the ones that are controversial.
More specifically, we obtain daily trending hashtags (both US and worldwide) on the platform for a period of three months (June 25 -- September 19, 2015).
Then, we obtain all tweets that use these hashtags, and create retweet graphs (as described in Section~\ref{sec:graph_building}).
Finally, we apply the \rwc measure on these conversation graphs to identify controversial hashtags.

The results can be explored in our online demo~\cite{garimella2016exploring}.\footnote{\url{https://users.ics.aalto.fi/kiran/controversy/table.php}.}
To mention a few examples, our system was able to identify the following controversial hashtags:
\begin{squishlist}
	\item \#whosiburningblackchurches (score \num{0.332}): A hashtag about the burning of predominantly black churches.\footnote{\url{https://erlc.com/article/explainer-whoisburningblackchurches}.}
	\item \#communityshield (score \num{0.314}): Discussion between the fans of two sides of a soccer game.\footnote{\url{https://en.wikipedia.org/wiki/2015_FA_Community_Shield}.}
	\item \#nationalfriedchickenday (score \num{0.393}): A debate between meat lovers and vegetarians about the ethics of eating meat.
\end{squishlist}
% Showing that our method can work on detecting controversial hashtags even in the wild, with no specific training data. %% there was some training involved, no? --michael

\begin{figure}
\centering
\includegraphics[height=0.5\textwidth, clip=true, trim=0 50 10 20]{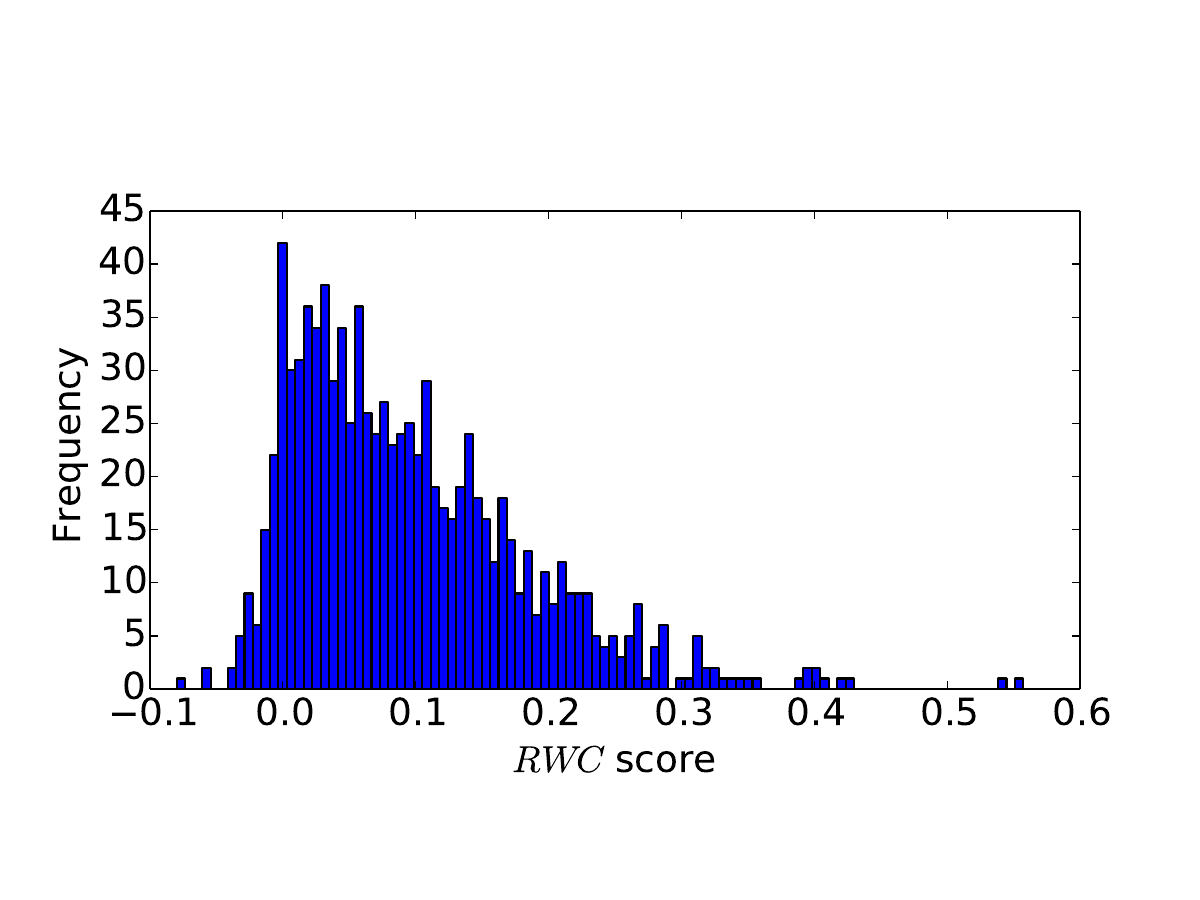}
\caption{Frequency of \rwc scores for hashtags trending from June to September 2015.}
\label{fig:RWC_frequency}
\vspace{-\baselineskip}
\end{figure}

Moreover, based on our experience with our system, most hashtags that are reported as trending on Twitter concern topics that are not controversial.
Figure~\ref{fig:RWC_frequency} shows the histogram of the \rwc score over the 924 trending hashtags we collected.
A majority of these hashtags have an \rwc score around zero.

% Mention something about the drawback of collecting trending hashtags (most of them are non-controversial).
%\note[michael]{Do/can we have a plot with the distribution of \rwc over all the in-the-wild hashtags? Besides saying that controversial hashtags are rare, we can also say whether they are easy to distinguish.}

%} % end color = red.

%%%%%%%%%%%%%%%%%%%%%%%%%%%%%%%%%%%%%%%%%%%%%%%%%%%%%%%%%%%%%%%%%%

%%%%%%%%%%%%%%%%%%%%%%%%%%%%%%%%%%%%%%%%%%%%%%%%%%%%%%%%%%%%%%%%%%

\section{Content}
\label{sec:content}

% \todo{Given that content seem not to work, is there any signal in content at all?}

In this section we explore alternative approaches to measuring controversy that use only the content of the discussion rather than the structure of user interactions.
As such, these methods do not fit in the pipeline described in Section~\ref{sec:pipeline}.
The question we address is ``does content help in measuring the controversy of a topic?''
In particular, we test two types of features extracted from the content.
The first is a typical IR-inspired bag-of-words representation.
The second involves sentiment-related features, extracted with NLP tools.

\subsection{Bag of words}
We take as input the raw content of the social media posts -- in our case, the tweets pertaining to a specific topic.
We represent each tweet as a vector in a high-dimensional space composed of the words used in the whole topic, after standard preprocessing used in IR (lowercasing, stopword removal, stemming).
Following the lines of our main pipeline, we group these vectors in two clusters %in order to try to find the two possible sides of a controversy.
by using CLUTO~\cite{karypis2002cluto} with cosine distance. % to cluster the data points.
% \note[gdfm]{which distance function do we use?. Cosine}

The underlying assumption is that the two sides, while sharing the use of the hashtag for the topic, use different vocabularies in reference to the issue at hand.
For example, for \#beefban a side may be calling for ``freedom'' while the opposing one for ``respect.''
%However, as we see from Figure/Table~\ref{fig:kldivergence}, this is not the case.
% We compute the KL divergence between the clusters in both controversial and non-controversial topics.
We use KL divergence as a measure of distance between the vocabularies of the two clusters, % in each topic.
%In addition we also compute 
and the I2 measure~\cite{maulik2002performance} of clustering heterogeneity.
% \note[gdfm]{Couldn't find the definition of I2 in the paper. -- Its on Page2, called the index I.}

We use an unpaired Wilcoxon rank-sum test at the $p=0.05$ significance level, but we are unable to reject the null hypothesis that there is no difference in these measures between the controversial and non-controversial topics.
Therefore, there is not enough signal in the content representation to discern between controversial and non-controversial topics with confidence.
This result suggests that the bag-of-words representation of content is not a good basis for our task.
It also agrees with our earlier attempts to use content to build the graph used in the pipeline (see Section~\ref{sec:graph_building}) -- which suggests that using content for the task of quantifying controversy might not be straightforward.

\subsection{Sentiment analysis}
Next, we resort to NLP techniques for sentiment analysis to analyze the content of the discussion.
We use SentiStrength~\cite{thelwall2013heart} trained on tweets to give a sentiment score in $[-4, 4]$ to each tweet for a given topic.
% \todo[gdfm]{Reference to the tool. Is it SentiStength?. Yes}
In this case we do not try to cluster tweets by their sentiment.
Rather, we analyze the difference in distribution of sentiment between controversial and non-controversial topics.

\begin{figure}[t]
\centering
\includegraphics[width=0.5\textwidth, clip=true, trim=0 40 10 50]{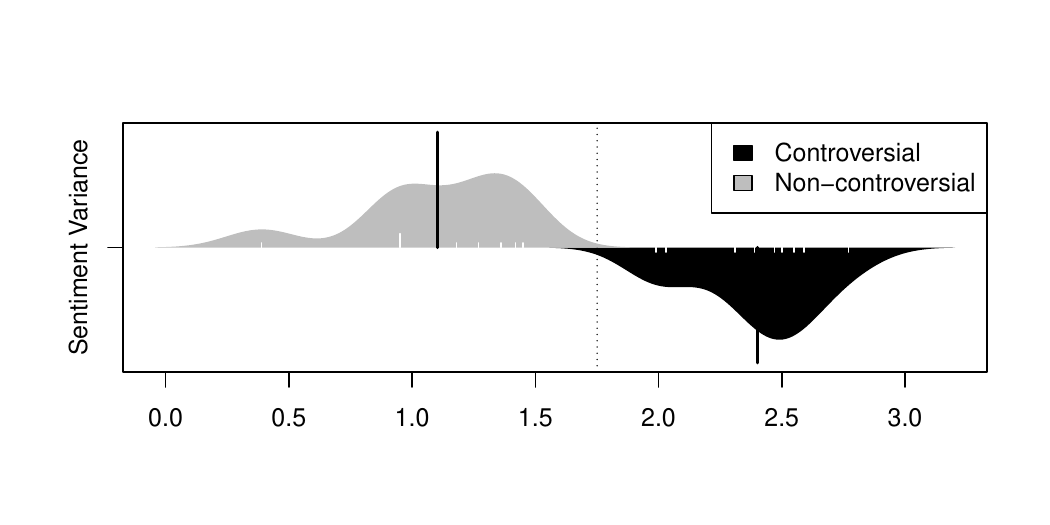}
\caption{Sentiment variance controversy score for controversial and non-controversial topics.}
\label{fig:sentiment}
%\vspace{-\baselineskip}
\end{figure}

While it is not possible to say that controversial topics are more positive or negative than non-controversial ones, we can detect a difference in their variance.
Indeed, controversial topics have a higher variance than non-controversial ones, as shown in Figure~\ref{fig:sentiment}.
Controversial ones have a variance of at least $2$, while non-controversial ones have a variance of at most $1.5$.
% \note[gdfm]{What about SPID? And given that the range is 4 (-2,2), can we normalize it to [0,1]? -- SPID is a bit difficult to compare because the mean could be -ve or +ve; The range is (-4,4). So normalizing might minimize the `separation'.}
% \note[gdfm]{OK, forget the SPID, can we instead normalize the range before? (divide by 4 the score). This way the expected value is zero and the maximum variance is 1. We can also use std. rather than variance to make it comparable with the sentiment value. It will decrease the range but I think it's not a problem.}

In practice, the ``tones'' with which controversial topics are debated are stronger, and sentiment analysis is able to detect this.
While this signal is clear, it is not straightforward to incorporate it into the measures based on graph structure.
Moreover, this feature relies on technologies that do not work reliably for languages other than English and hence cannot be applied for topics such as \#russia\_march.

%Signal present, not clear how to incorporate it into the network.
%Not trivial to generalize for languages other than english. Refer to Table~\ref{tab:sentiment}.

%Future work.

%\begin{table}
%\centering
%\caption{\label{tab:sentiment}Distribution of sentiment on various datasets.}
%\begin{tabular}{l r r}
%\toprule
%Topic & variance & normalized Var \\
%\midrule
%\#beefban & 2.55 & 0.32 \\
%\#nemtsov & 1.99 & 0.25 \\
%\#netanyahuspeech & 2.77 & 0.35 \\
%\#indiasdaughter & 2.47 & 0.31 \\
%\#baltimoreriots & 2.59 & 0.32 \\
%\#indiana & 2.5 & 0.31 \\
%\#ukraine & 2.31 & 0.29 \\
%\#gunsense & 2.03 & 0.25 \\
%\#leadersdebate & 2.39 & 0.29 \\
%\midrule
%\#sxsw & 1.18 & 0.15 \\
%\#1dfamheretostay & 0.95 & 0.12 \\
%\#germanwings & 0.39 & 0.05 \\
%\#mothersday & 1.45 & 0.18 \\
%\#nepal & 1.36 & 0.17 \\
%\#ultralive & 0.95 & 0.12 \\
%\#FF & 0.95 & 0.12 \\
%\#jurassicworld & 1.27 & 0.16 \\
%\#nationalkissingday & 1.42 & 0.17 \\
%\bottomrule
%\end{tabular}
%\todo{Transform into a Figure: horizontal bean plot}
%\end{table}

%%%%%%%%%%%%%%%%%%%%%%%%%%%%%%%%%%%%%%%%%%%%%%%%%%%%%%%%%%%%%%%%%%

%%%%%%%%%%%%%%%%%%%%%%%%%%%%%%%%%%%%%%%%%%%%%%%%%%%%%%%%%%%%%%%%%%

\section{Discussion}
\label{sec:discussion}
The task we tackle in this work is certainly not an easy one, and this study has some limitations, which we discuss in this section.
We also report a set of negative results that we produced while coming up with the measures presented.
We believe these results will be very useful in steering this research topic towards a fruitful direction.
Table~\ref{tab:summary} provides a summary of the various graph building strategies and controversy measures we tried for quantifying controversy.

%\note[AG]{Perhaps we can summarize all the options we tried
%(graphs, measures, content, partitioning methods, etc.) in a table, 
%grouped by type. It could also go in the discussions section.}

\begin{table}[]
\centering
\caption{Summary of various graph building and controversy measures tried. Methods that worked reliably are marked in bold.}
\label{tab:summary}
\begin{tabular}{l | r}
\toprule
\multirow{5}{*}{Graphs}   & \textbf{Retweet}                                       \\
                          & \textbf{Follow}                                        \\
                          & Content                                       \\
                          & Mention                                       \\
                          & Hybrid (content + retweet, mention + retweet) \\
\midrule
\multirow{9}{*}{Measures} & \textbf{Random Walk}                                   \\
                           & \textbf{Edge Betweenness}                              \\
                           & Embedding                                     \\
                           & Boundary Connectivity                         \\
                           & Dipole Moment                                \\
			  & Cut-based Measures (conductance, cut ratio) \\
			  & \textbf{Sentiment Analysis} \\
			  & Modularity \\
			  & SPID \\
\bottomrule
\end{tabular}
\end{table}

\subsection{Limitations}

\spara{Twitter only.}
We present our findings mostly on datasets coming from Twitter.
While this is certainly a limitation, Twitter is one of the main venues for online public discussion, and one of the few for which data is available.
Hence, Twitter is a natural choice.
In addition, our measures generalize well to datasets from other social media and the Web.

\spara{Choice of data.}
We manually pick the controversial topics in our dataset, which might introduce bias.
In our choice we represent a broad set of typical controversial issues coming from religious, societal, racial, and political domains.
Unfortunately, ground truths for controversial topics are hard to find, especially for ephemeral issues.
However, the topics are unanimously judged controversial by the authors.
Moreover, the hashtags represent the intuitive notion of controversy that we strive to capture, so human judgement is an important ingredient we want to use.
%While it would be theoretically possible to crowdsource the labeling of the hashtags, designing the task has its own challenges, and we deem it too complicated to obtain a reliable result.

\spara{Overfitting.}
While this work presents the largest systematic study on controversy in social media so far, we use only 20 topics for our main experiment.
Given the small number of examples, the risk of overfitting our measures to the dataset is real.
We reduce this risk by using only $40\%$ of the topics during the development of the measures.
Additionally, our measures agree with previous independent results on external datasets, which further decreases the likelihood of overfitting.

\spara{Reliance on graph partitioning.}
Our pipeline relies on a graph partitioning stage, whose quality is fundamental for the proper functioning of the controversy measures.
Given that graph partitioning is a hard but well studied problem, we rely on off-the-shelf techniques for this step.
A measure that bypasses this step entirely is highly desirable, and we report a few unsuccessful attempts in the next subsection.
%However, we find that METIS is fairly robust and its output is adequate in most cases.

\spara{Multisided controversies.}
Not all controversies involve only two sides with opposing views.
Some times discussions are multifaceted, or there are three or more competing views on the field.
The principles behind our measures neatly generalize to multisided controversies.
However, in this case the graph partitioning component needs to automatically find the optimal number of partitions.
%Moreover, these kinds of controversies are inherently less common and thus harder to study systematically.
We defer experimental study of such cases to an extended version of this paper.

\spara{Evaluation.}
Defining what is controversial/polarized can be subjective. There are many ways to define what is controversial, depending on the context, subject and field of study, e.g. See~\cite{bramson2016disambiguation} for around a dozen ways to define polarization.
Our evaluation is based on our intuitive labelling that a topic is controversial/polarized. This might not always be true, but given that the alternative is to hand-label/survey the thousands of users, we presume that this assumption is reasonable for developing methods that can be adapted to large scale systems.

\subsection{Negative results}
We briefly review a list of methods that failed to produce reliable results and were discarded early in the process of refining our controversy measures.

\spara{Mentions graph.}
\citet{conover2011political} rely on the mention graph in Twitter to detect controversies.
However, in our dataset the mention graphs are extremely sparse given that we focus on short-lived events.
Merging the mentions into the retweet graph does not provide any noticeable improvement.

Previous studies have also shown that people retweet similar ideologies but mention across ideologies~\cite{bessi2014social}.
We exploit this intuition by using correlation clustering for graph partitioning, with negative edges for mentions.
Alas, the results are qualitatively worse than those obtained by METIS.
%While it is possible for a more sophisticated METIS-like algorithm to extract a useful signal from the mentions graph, its size is negligible compared to the retweet graph, so it is not immediately obvious how much it should influence the output.

\spara{Cuts.}
Simple measures such as size of the cut of the partitions do not generalize across different graphs.
Conductance (in all its variants) also yields poor results.
Prior work identifies controversies by comparing the structure of the graph with randomly permuted ones~\cite{conover2011political}.
Unfortunately, we obtain equally poor results by using the difference in conductance with cuts obtained by METIS and by random partitions.

\spara{Community structure.}
Good community structure in the conversation graph is often understood as a sign that the graph is polarized or controversial.
However, this is not always the case.
We find that both assortativity and modularity (which have been previously used to identify controversy) do not correlate with the controversy scores, and are not good predictors for how controversial a topic is.
The work by~\citet{guerra2013measure} presents clear arguments and examples of why modularity should be avoided.

\spara{Partitioning.}
As already mentioned, bypassing the graph partitioning to compute the measure is desirable.
We explore the use of the all pairs expected hitting time computed by using SimRank~\cite{jeh2002simrank}.
We compute the SPID (ratio of variance to mean) of this distribution, however results are mixed.

\subsection{Conclusions}
% Quantifying controversy in social media is a challenging task, 
% with important implications for the health of online discussions and the formation of public opinion.
In this paper, we performed the first large-scale systematic study for quantifying controversy in social media.
We have shown that previously-used measures are not reliable and demonstrated that controversy can be identified both in the retweet and topic-induced follow graph.
We have also shown that simple content-based representations do not work in general, 
while sentiment analysis offers promising results.

% To tackle the controversy-quantification challenge we have proposed a three-stage pipeline,
% which allows to measure controversy in an unsupervised fashion when deployed in-the-wild on the field.
% We have developed and tested several measures for the last stage of our pipeline.
Among the measures we studied, the random-walk-based $\mathit{RWC}$ most neatly 
separates controversial topics from non-controversial ones.
Besides, our measures gracefully generalize to datasets from other domains and previous studies.

This work opens several avenues for future research.
First, it is worth exploring alternative approaches and testing additional features, 
such as, following a generative-model-based approach, 
or exploiting the temporal evolution of the discussion of a topic~\citep{garimella2017effect}.

From the application point of view, 
the controversy score can be used to generate recommendations 
that foster a healthier ``news diet'' on social media.
Given the ever increasing impact of polarizing figures in our daily politics and the rise in polarization in the society~\cite{dimock2014political,garimella2017long}, it is important to not restrict ourselves to our own `bubbles' or `echo chambers'~\cite{pariser2011filter,sunstein2009republic}.
Our methods for identifying controversial topics can be used as building blocks for designing such systems to reduce controversy on social media~\cite{garimella2017reducing, garimella2017factors} by connecting social media users with content outside their own bubbles.

Finally, polarization by itself may not be a wholly negative phenomenon.
Several studies~\cite{mutz2002consequences,dahlberg2007rethinking} argue that a democracy needs deliberation, and polarization enable such a deliberation to happen in the public, to a certain extent, thus informing people about the issues and arguments from different sides.
Given such a setting, it is of paramount importance to understand to what extent a discussion is polarized, so that things do not spiral out of control, or create isolated echo chambers. 
Our paper contributes methods that are useful in this setting, and enable measuring the degree of polarization of a topic in a domain-agnostic fashion.

%%%%%%%%%%%%%%%%%%%%%%%%%%%%%%%%%%%%%%%%%%%%%%%%%%%%%%%%%%%%%%%%%%

\spara{Acknowledgements.}
This work has been supported by the Academy of Finland project ``Nestor'' (286211) and the EC H2020 RIA project ``SoBigData'' (654024).

\bibliographystyle{ACM-Reference-Format-Journals}
\bibliography{biblio}

\end{document}